\documentclass[twocolumn]{openjournal}

\usepackage{orcidlink}
\usepackage{lineno}
\usepackage{comment}
\usepackage{color}
\usepackage{hyperref}
\hypersetup{
	colorlinks=true,
	urlcolor=blue,    
	linkcolor=black,  
	citecolor=blue,   
}
\newcommand{\orcidauthor}[3]{\author{\href{http://orcid.org/#1}{#2$^{#3}$}}}

\usepackage{amsmath}

\shorttitle{Model for Transients Preceding Binary Mergers}
\shortauthors{Tsuna et al.}

\begin{document}

\title{Semi-analytical Light Curve Model for Transients Preceding Binary Mergers. I: Supernova Precursor Emission from Compact Object Companions\vspace{-1.5cm}}

\email{daichi.tsuna@cfa.harvard.edu}

\orcidauthor{0000-0002-6347-3089}{Daichi Tsuna}{1}

\orcidauthor{0000-0002-1417-8024}{Morgan Macleod}{1}

 \orcidauthor{0000-0002-5814-4061}{V. Ashley Villar}{1,2}
 \affiliation{$^{1}$Center for Astrophysics $|$ Harvard \& Smithsonian, 60 Garden St, Cambridge, MA 02138, USA}
 \affiliation{$^{2}$The NSF AI Institute for Artificial Intelligence and Fundamental Interactions}

\begin{abstract}
A binary undergoing dynamically unstable mass transfer could rapidly shrink its orbit and lead to a merger, with the whole process often observable in human timescales. We construct a semi-analytical light curve model of binary systems composed of a star and a compact object accretor, that would display long-rising accretion-powered emission prior to a final merger-driven explosion. We apply the model to the precursors of interacting supernovae (SNe) that display long-rising light curves of years, SN 2023zkd, 2023fyq and 2021qqp, demonstrating the model's capability of inferring the mass-transfer history and constraining the progenitor binary system. The model and the parameter inference framework, encapsulated in a publicly-released script, can be applied to existing long-rising precursors of SNe as well as many SN precursors to be discovered by surveys like those from the Vera C. Rubin Observatory. 
\end{abstract}


\section{Introduction}
Stars more massive than our Sun typically live in binary or multiple systems, often with close separations such that they interact through their lives \citep{Sana12,Duchene13,DeMarco17,Sana25}. Mass transfer occurs in such close binaries when the donor star expands and overfills its Roche lobe. This has broad consequences on the binary's orbit and stability, depending on the donor's structure, binary's mass ratio, and the mass and angular momentum lost from the system \citep[e.g.,][]{Hjellming87,Soberman97}. 

In some cases mass transfer can be dynamically unstable, eventually leading to merger of the two stars followed by a common-envelope phase \citep[e.g.,][]{Paczynski76,Ivanova13}. Such stellar mergers have been associated with a class of transients called luminous red novae (LRNe). The particular case was the Galactic event V1309 Sco \citep{Tylenda11}, where the light curve transitioned from that of an eclipsing binary to a ramp-up lasting for years. This behavior is now explained by increasing orbital decay and mass ejection due to nonconservative mass transfer \citep{Pejcha14,Pejcha17,Macleod18a,Macleod20}. 

More recently, there is an increasing number of supernovae (SNe) where archival data revealed slowly-rising precursors of luminosities of $10^{40}$--$10^{42}$ erg s$^{-1}$ lasting for years, with evolution reminiscent of LRNe \citep{Hiramatsu24,Dong24,Gagliano25b}. If interpreted as a one-off mass ejection event, the required mass to explain the rise time by diffusion is unphysically huge \citep[$>100~M_\odot$,  e.g. Figure 5 of][]{Khatami24}, suggesting instead a continuous energy injection that gradually ramps up with time. In analogy to LRNe, such precursor events were theoretically suggested as unstable mass transfer with a compact object accretor, with the final explosion plausibly triggered by the merger of the two stars \citep{Tsuna24,Dong24}. In one event SN 2023fyq, the unstable mass transfer scenario has also been supported by independent radio observations, which probe circumstellar material (CSM) emitted years before the final explosion \citep{Baer-Way25}. These events are therefore hinting at the existence of a novel population of merger-driven SN-like explosions that are distinct from traditional explosions triggered by core-collapse of massive stars.

The mass transfer process creates circumbinary material with varying mass and extent that the final explosion due to merger eventually interacts with. When looking at only the final explosion, such merger-driven explosions could be mistaken as interacting ``core-collapse" SNe of e.g. Type IIn/Ibn \citep{Chevalier12,Metzger22,Tsuna24}, as the energetics of merger-driven explosions can be comparable to SNe \citep{Zhang01,Soker19,Schroder20}. Detections of precursors (or lack thereof) can be an indispensable way to distinguish between these two explosion mechanisms.

The slow-rising, dim nature of these precursors make them ideal targets for Vera C. Rubin Observatory’s Legacy Survey of Space and Time \citep[LSST;][]{Ivezic19}. In the Rubin era we expect to find tens to hundreds of precursors of interacting SNe per year \citep[e.g.,][]{Strotjohann24,Gagliano25a}, and there are ongoing efforts to efficiently find these events \citep{Dong25}. We expect that a significant fraction will also be detected with long-rising light curves anticipated from binary mergers. While such precursors carry important information of the binary system before merger, a model framework to better understand this has been lacking.

Motivated by these prospects, we construct a simple semi-analytical light curve model of events powered by unstable mass transfer onto an accreting compact companion, connecting these to the long-rising precursors of interacting SNe. We also present a parameter inference framework for extracting the binary mass transfer history from the observed light curves, which can give unique constraints on the progenitor binary system.\footnote{Our light curve model and the parameter inference for example precursor events in this work are publicly available in: \url{https://github.com/DTsuna/merger-light-curve-models}.}

Section \ref{sec:model} shows our model for calculating the luminosity and temperature of the precursor emission. Section \ref{sec:applications} discusses the expected range of the model parameters from binary models. Section \ref{sec:fitting_events} introduces the parameter inference framework using our model, with applications to observed precursor events of interacting SNe. We conclude in Section \ref{sec:conclusion} with directions for future work.

\section{Semi-analytical Model}
\label{sec:model}
Our work focuses on the case of binary systems with a neutron star (NS) or black hole (BH) accretor, and we define the masses of the donor and accretor as $M_*$ and $M_\bullet$ respectively. Mass transfer is expected to be unstable for binaries with highly unequal mass ratios, though the exact mass ratios are under debate and likely vary with donor and accretor properties \citep[$M_\bullet/M_*\lesssim 0.2$--$0.6$; e.g.,][]{Pavlovskii15, Henneco24, Ercolino24}. For a compact object accretor, we thus typically expect the donor to be a massive star, most likely to have evolved off from main sequence and initiated mass transfer due to rapid radial expansion.

In the following, we consider the expected case where mass transfer is highly non-conservative, and the compact object accretor launches a quasi-steady, optically thick wind that reprocesses its accretion power. We solve the luminosity and temperature of the wind-reprocessed emission following the framework of \cite{Piro20}, also accounting for reddening due to dust formation.

\subsection{Mass Transfer Evolution}
In binaries undergoing unstable mass transfer, the donor's increasing Roche lobe overflow with time leads to a runaway evolution in orbit and mass transfer rate, until the accretor is engulfed by the donor's envelope. In this phase, we expect the mass transfer rate to evolve as \citep{Webbink77,Pejcha14}
\begin{eqnarray}
    |\dot{M}_{*, \rm dyn}(t)| \propto (t_m-t)^{-\delta},
    \label{eq:Mdot_formula_dyn}
\end{eqnarray}
where $\delta>0$ is a constant, and $t=t_m$ denotes the time of merger when $\dot{M}_{*, \rm dyn}$ diverges. The power-law form is motivated by the relation of the mass transfer rate from Roche-lobe overflow \citep{Paczynski72,Macleod20},
\begin{eqnarray}
  \dot{M}_{*, \rm dyn}\propto -\left(\frac{R_*-R_L}{R_L}\right)^{n+3/2},
  \label{eq:Mdot_vs_RLOF}
\end{eqnarray}
where $R_*$, $R_L$ are respectively the radius of the donor and its Roche lobe, $n$ is the polytropic index of the donor's envelope, and the negative sign indicates mass loss from the donor. For the early phase of Roche-lobe overflow where $R_*-R_L\ll R_L$, we expect both $d\ln R_*/d\ln M_*$ (the adiabatic mass-radius relation) and $d\ln R_L/d\ln M_*$ to be nearly constant over a small change in mass \citep{Webbink77}, and thus to first order
\begin{eqnarray}
    \left(\frac{R_*-R_L}{R_L}\right)^{n+3/2} \propto \left(\frac{M_{*,0}-M_*}{M_{*,0}}\right)^{n+3/2},
    \label{eq:RLOF_vs_Mstar}
\end{eqnarray}
where $M_{*,0}$ is the initial mass of the donor. We then find from solving the differential equation for ${M}_*(t)$ from equations (\ref{eq:Mdot_vs_RLOF}) and (\ref{eq:RLOF_vs_Mstar}),
\begin{eqnarray}
    \dot{M}_{*, \rm dyn} &\propto& -\left(M_{*,0}-M_*\right)^{n+3/2}, \nonumber \\
    &\propto& \left[(t_C-t)^{\!-2/(2n+1)}\right]^{\!n+3/2} \propto (t_C-t)^{\!-(2n+3)/(2n+1)}, \nonumber
\end{eqnarray}
where $t_C$ is the constant of integration. Comparing this with equation (\ref{eq:Mdot_formula_dyn}), we expect $t_C=t_m$ and the relation
\begin{eqnarray}
    \delta = 1+\frac{2}{2n+1},
\end{eqnarray}
with representative values of $\delta \approx 1.5~(1.286)$ for convective, gas-pressure dominated (radiative, radiation-dominated) envelopes with $n\approx 3/2~(3)$. For a polytrope, a finite stellar radius implies $n<5$, and we expect a lower limit of $\delta>13/11(\approx 1.182)$. As $\delta>1$, the donor mass is dominantly lost closer to merger as mass transfer enters a runaway. 

The assumptions for equation (\ref{eq:Mdot_formula_dyn}) are less justified as the Roche-filling fraction $(R_*-R_L)/R_L$ approaches unity. This occurs roughly when the time before merger becomes comparable to the donor's dynamical timescale \citep[e.g., Figure 19 of][]{Macleod18a}
\begin{eqnarray}
    t_{\rm dyn}\approx \sqrt{\frac{R_*^3}{GM_*}}\sim  3\ {\rm day}\left(\frac{R_*}{100\ R_\odot}\right)^{\!\!3/2}\left(\frac{M_*}{30\ M_\odot}\right)^{\!\!-1/2},
\end{eqnarray}
where $G$ is the gravitational constant, and we adopted values for $R_*, M_*$ representative of inflated donors that are promising progenitors for stellar mergers. Hence equation (\ref{eq:Mdot_formula_dyn}) is good approximation at $\gtrsim$ weeks before merger for donors around the Hertzsprung gap, and at $\gtrsim$ several months before merger for red supergiant donors.

The dynamical phase could be preceded by mass transfer with a much longer timescale (e.g. thermal/nuclear timescale), that evolves much more slowly over time \citep[e.g.,][]{Hjellming87,Ge15,Temmink25}. To approximately capture this effect, we adopt the following form for the mass transfer rate
\begin{eqnarray}
|\dot{M}_*(t)| =  \dot{M}_0\left[1+\left(\frac{t_m-t_0}{t_m-t}\right)^{\!\!\delta}\right]\  (t_{\rm MT}<t<t_m),
     \label{eq:Mdot_formula}
 \end{eqnarray} 
 where $t=t_{\rm MT}(\leq t_0)$ is the onset of mass transfer, and $t=t_0$ is the onset of the dynamically unstable phase when $\dot{M}_*$ starts to ramp up. This prescription gives $|\dot{M}_*(t)|\approx \dot{M}_0\ ({\rm const.})$ well before the dynamical phase ($t_m-t \gg t_m-t_0$), which transitions to the desired $|\dot{M}_*(t)|\propto (t_m-t)^{-\delta}$ when mass transfer enters the dynamical phase ($t_m-t \lesssim t_m-t_0$).

\subsection{Formation of a Wind}
Before merger, the donor transfers a significant fraction $f\approx 0.25q$ of its mass over the star's dynamical timescale, with $f$ mostly scaling with the binary mass ratio $q=M_\bullet/M_*$ \citep{Macleod20b}. Approximating the mass transfer rate over the last dynamical time as $|\dot{M}_*|\approx fM_*/t_{\rm dyn}$ at $t=t_m-t_{\rm dyn}$, the mass-transfer rates at many dynamical times before merger ($t_m-t\gg t_{\rm dyn}$) is roughly expressed by the donor's properties as
\begin{eqnarray}
    |\dot{M}_*|&\approx& \left(\frac{t_m-t}{t_{\rm dyn}}\right)^{-\delta}\frac{fM_*}{t_{\rm dyn}}\nonumber \\
    &\sim& 0.3\ M_\odot~{\rm yr}^{-1}\nonumber\\
    &\times& \left(\frac{q}{0.25}\right)\left(\frac{t_m-t}{1~ \rm yr}\right)^{\!\!-1.4}\left(\frac{R_*}{100\ R_\odot}\right)^{\!\!0.6}\left(\frac{M_*}{30~M_\odot}\right)^{\!\!0.8}.
    \label{eq:Mdot_estimate}
\end{eqnarray}
Here we adopted $\delta\approx 1.4$ in the scaling, and variations of $\delta\approx 1.3$--$1.5$ change this by $\lesssim 50~\%$ for a broad range of $R_*\approx 10$--$1000~R_\odot$. The mass transfer rate is orders of magnitude larger than the standard Eddington-limited accretion rate of stellar-mass compact objects, $\dot{M}_{\rm Edd}\sim 2\times 10^{-7}~M_\odot\ {\rm yr}^{-1}(M_\bullet/10~M_\odot)$. 

The accretor is generally unable to fully accept such an intense mass transfer from the donor, and we expect outflow of material through the L2 point and/or by a disk wind from the vicinity of the accretor \citep{Lu23,Scherbak25,Scherbak26}. 
Non-conservative mass transfer is usually prescribed by a fraction $\beta (\leq 1)$ being transferred to the accretor, and a mass loss rate of $\dot{M}_w\approx (1-\beta)|\dot{M}_*|$ being lost from the binary as some form of a ``wind". Given the above high $|\dot{M}_*|$ of our interest, we adopt $\beta\ll 1$, leading to
\begin{eqnarray}
    \dot{M}_w \approx |\dot{M}_*| \approx \dot{M}_0\left[1+\left(\frac{t_m-t_0}{t_m-t}\right)^{\delta}~\right].
    \label{eq:Mdot_wind}
\end{eqnarray}

We assume the wind is launched quasi-spherically from a characteristic radius $r_{\rm in}$, carrying an accretion power parameterized by an efficiency $\epsilon_{\rm acc}$ as
\begin{eqnarray}
    L_{\rm acc} &=& \epsilon_{\rm acc}\dot{M}_w c^2 \nonumber \\
    &\sim& 6\times 10^{41}\ {\rm erg\ s^{-1}}\left(\frac{\epsilon_{\rm acc}}{10^{-4}}\right)\left(\frac{\dot{M}_w}{10^{-1}\ M_\odot\ {\rm yr}^{-1}}\right),
\end{eqnarray}
where $c$ is the speed of light. We regard the radius $r_{\rm in}$ as the radii where rough equipartition is realized between kinetic energy and internal energy, with the latter being dominated by radiation for our cases of interest. 

As mass transfer is highly non-conservative and most of the mass is not accreted onto the compact object, we expect $\epsilon_{\rm acc}\ll 1$ with some dependence on $|\dot{M}_*|$. We approximate this as a power-law evolution with $|\dot{M}_*|$,
\begin{eqnarray}
    \epsilon_{\rm acc} = \epsilon_{\rm acc,0}\left(\frac{|\dot{M}_*|}{\dot{M}_0}\right)^{n_{\rm acc}},
    \label{eq:eps_evolution}
\end{eqnarray}
where $\epsilon_{\rm acc,0}$ is the efficiency for $|\dot{M}_*|=\dot{M}_0$, and $n_{\rm acc}$ is the power-law index we assume as constant ($\leq 0$).
For a compact object companion we generally expect the wind launching to mainly originate from (super-Eddington) accretion \citep[e.g.,][]{Shakura73,Ohsuga05,Jiang14,Sadowski14}, while our light curve model remains agnostic to the precise mechanisms and aims to derive these values from the observed light curves. We nevertheless discuss the expected values of these model parameters in Section \ref{sec:applications}, where we adopt an analytic prescription for super-Eddington accretion. 

\subsection{Emission from the Wind: Luminosity}
\begin{figure*}
    \centering
 \includegraphics[width=\linewidth]{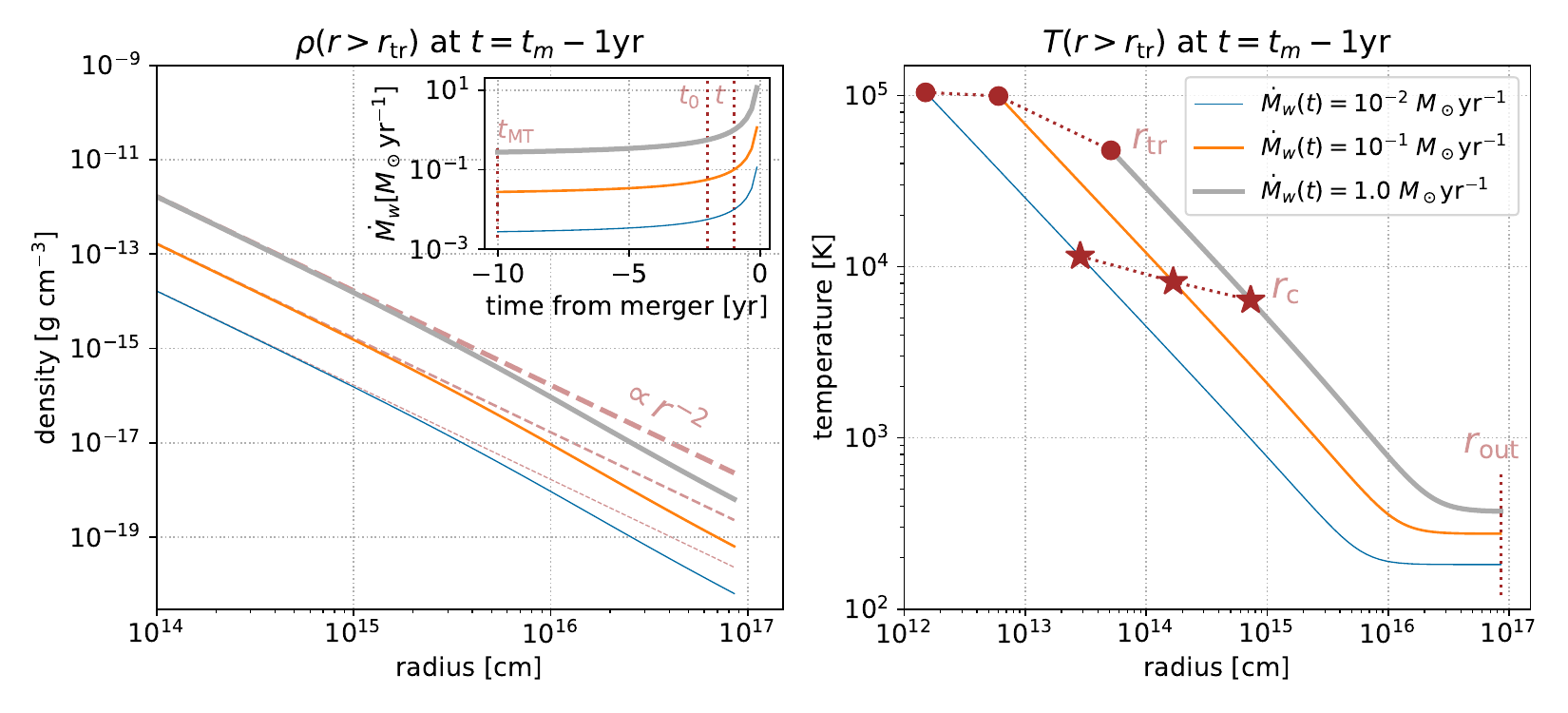}
    \caption{Example density (left) and temperature (right) profiles of the wind, for three cases of $\dot{M}_w$ at 1 year before merger. Due to the increasing $\dot{M}_w$ in the dynamical phase (inset of left panel), the outer part of the wind steepens from the standard $r^{-2}$ density profile, shown as dashed lines in the left panel. Dotted lines on the right panel show the wind's outermost radius $r_{\rm out}$, trapping radius $r_{\rm tr}$ (dots), and the color radius $r_c$ (stars) where the emission temperature $T_{\rm rad}$ is set. The wind is assumed to have a solar-like composition, and the following parameters are adopted: a constant $\epsilon_{\rm acc}=10^{-4}$ ($v_w\approx 3000\ {\rm km\ s^{-1}}, n_{\rm acc}=0$), $\delta=1.4, r_{\rm in}=10^{12}$ cm, and $t_{\rm MT}, t_0$ respectively being 10, 2 years before merger.}
    \label{fig:rho_T_profile}
\end{figure*}

A significant fraction of the accretion power is carried by radiation at $r_{\rm in}$, but the wind can initially be  optically thick and expand nearly adiabatically. As the wind expands, this radiation energy is degraded as it is converted to kinetic energy via $PdV$ work. We aim to develop a light curve model capturing these effects. 

We divide the wind into multiple ``shells" launched at times $\tilde{t}$. Let us consider a shell of wind material launched at time $\tilde{t}$ over a time $\Delta t$, with the accretion power split equally between kinetic and internal energy upon launch as $L_{\rm acc}\Delta t/2$. The initial internal energy is
\begin{eqnarray}
    E_{\rm int, 0} \approx \frac{1}{2}L_{\rm acc} \Delta t= \frac{1}{2}\epsilon_{\rm acc}(\dot{M}_w\Delta t) c^2,
\end{eqnarray}
and the wind velocity $v_w$ is obtained from the kinetic energy $L_{\rm acc}\Delta t/2\approx (\dot{M}_w\Delta t) v_w^2/2$ as
\begin{eqnarray}
v_w&\approx& \sqrt{2\times \frac{L_{\rm acc}\Delta t/2}{\dot{M}_w\Delta t}} \nonumber \\
&=&(\sqrt{\epsilon_{\rm acc}})c \sim 3000\ {\rm km\ s^{-1}}\left(\frac{\epsilon_{\rm acc}}{10^{-4}}\right)^{\!1/2}.
\end{eqnarray}
At time $t'$ after launch, the shell increases its volume as $V_{\rm sh}=4\pi[r_{\rm in}+(v_w t')]^2(v_w\Delta t)$, and its internal energy $E_{\rm int}$ (assumed to be dominated by radiation) evolves as
\begin{eqnarray}
    \frac{dE_{\rm int}}{dt'} &=& -\frac{E_{\rm int}}{3V_{\rm sh}}\frac{dV_{\rm sh}}{dt'} - L_{\rm shell}\\
    &=& -\frac{2E_{\rm int}}{3(t'+r_{\rm in}/v_w)} - \frac{E_{\rm int}}{t_{\rm diff}},
\end{eqnarray}
where the first term is adiabatic cooling by $PdV$ work, and the second term is the radiative cooling of the shell regulated by the diffusion timescale through the wind
\begin{eqnarray}
    t_{\rm diff} &\approx& \frac{\kappa \dot{M}_w}{4\pi v_wc} \nonumber \\
    &\sim& 0.2\ {\rm day} \left(\frac{\kappa}{0.3\ {\rm cm^2\ g^{-1}}}\right)\nonumber \\
    &&\left(\frac{\dot{M}_w}{10^{-1}\ M_\odot\ {\rm yr}^{-1}}\right) \left(\frac{v_w}{3000\ {\rm km\ s^{-1}}}\right)^{\!\!-1},
    \label{eq:t_diff}
\end{eqnarray}
where $\kappa$ is the opacity, generally dominated by scattering opacity for the scenarios we consider. For a quasi-steady wind  in which $t_{\rm diff}$ is independent of $t'$, we obtain a solution to the above differential equation, with initial conditions of $E_{\rm int}=E_{\rm int,0},~r=r_{\rm in}$ at $t'=0$ as
\begin{eqnarray}
    E_{\rm int}(t') &=& E_{\rm int, 0}\left(\frac{r_{\rm in}}{r_{\rm in}+v_w t'}\right)^{2/3} \exp\left[-\frac{t'}{t_{\rm diff}}\right] , \\
    L_{\rm shell}(t') &=& \frac{E_{\rm int, 0}}{t_{\rm diff}}\left(\frac{r_{\rm in}}{r_{\rm in}+v_w t'}\right)^{2/3} \exp\left[-\frac{t'}{t_{\rm diff}}\right] .
\end{eqnarray}

For a quasi-steady wind, this indicates that the bulk of the radiation is emitted at $t'\sim t_{\rm diff}$ when the shell reaches the trapping radius \citep{Piro20},
\begin{eqnarray}
    r_{\rm tr} \approx r_{\rm in} + v_wt_{\rm diff}= r_{\rm in} + \frac{\kappa \dot{M}_w}{4\pi c}.
\end{eqnarray}

We can now integrate over all the shells with time to obtain the light curve as a function of $t$. Integrating over the shells launched at time $\tilde{t}~(0<\tilde{t}<t)$, we can sum up the contribution from each shell as
\begin{eqnarray}
    &&L_{\rm rad}(t) \nonumber\\
    &=&\int_0^t d\tilde{t} \frac{L_{\rm acc}(\tilde{t})}{2t_{\rm diff}(\tilde{t})} \left[\frac{r_{\rm in}}{r_{\rm in}+v_w (t-\tilde{t})}\right]^{\!2/3} \exp\left[-\frac{t-\tilde{t}}{t_{\rm diff}(\tilde{t})}\right], 
    \label{eq:L_rad_steady}
\end{eqnarray}
with the factor 2 in the denominator again due to equipartition at the launching radius. For a quasi-steady wind $L_{\rm acc}(\tilde{t})$ and $t_{\rm diff}(\tilde{t})$ are approximately constant. Then the exponential factor causes only the most recent few $t_{\rm diff}$ from $t$ to contribute, and we recover the relation $L_{\rm rad}\approx (L_{\rm acc}/2)(r_{\rm in}/r_{\rm tr})^{2/3}$ from \cite{Piro20} at differences within factors of order unity. Practically, this means that the integral is dominated by the contributions of the most recent diffusion times, i.e. contributions from shells at radii within a couple trapping radii. This gives a quasi-steady state luminosity, valid so long as $\dot{M}_w$ does not strongly evolve over $t_{\rm diff}$.

\subsection{Emission from the Wind: Temperature}
\label{sec:model_temp}
We calculate the emission temperature by solving the reprocessing by the wind. At a given time, the wind extends out to a radius
\begin{eqnarray}
    r_{\rm out}(t) = r_{\rm in} + (v_{w|t=t_{\rm MT}})(t-t_{\rm MT})
\end{eqnarray}
with a density profile
\begin{eqnarray}
    \rho(r,t) &=& \frac{\dot{M}_w}{4\pi r^2 v_w},
\end{eqnarray}
where we note that the values of $\dot{M}_w, v_w$ are those not at time $t$, but when the wind was launched ($t-r/v_w$). The profile is in general steeper than a steady wind $\rho\propto r^{-2}$, as larger $r$ traces mass loss in the more distant past with lower $\dot{M}_w$ and higher $v_w\propto \epsilon_{\rm acc}^{1/2}$ (equations \ref{eq:Mdot_formula}, \ref{eq:eps_evolution}).

The temperature profile outside the trapping radius is set by photons diffusing with roughly constant luminosity $L_{\rm rad}$ from equation (\ref{eq:L_rad_steady}). The temperature in the wind follows the diffusion approximation as \citep{Piro20}
\begin{eqnarray}
    L_{\rm rad} = - \frac{4\pi r^2 ac}{3\kappa \rho}\frac{\partial T^4}{\partial r} ,
    \label{eq:diffusion_approx}
\end{eqnarray}
where $a$ is the radiation constant. As an outer boundary condition, we assume that the outermost radius of the wind obeys the local radiative equilibrium temperature
\begin{eqnarray}
    T(r=r_{\rm out}) = \left(\frac{L_{\rm rad}}{4\pi r_{\rm out}^2 ac}\right)^{\!\!1/4}.
\end{eqnarray}
We note that radiative equilibrium is not well established at low densities, and the temperature at $r\approx r_{\rm out}$ could be higher than what is assumed here. However, the temperature profile $T(r)$ at the interior, obtained by integrating equation (\ref{eq:diffusion_approx}) outside-in, is nearly insensitive to the choice of $T(r=r_{\rm out})$ due to the steep $T^4$ dependence in equation (\ref{eq:diffusion_approx}).

Figure \ref{fig:rho_T_profile} shows examples of the density and temperature profiles of our wind model, for three cases of $\dot{M}_w$ at 1 year before merger of $10^{-2}$--$1~M_\odot\ {\rm yr}^{-1}$ expected for a range of donor properties (see equation \ref{eq:Mdot_estimate}). We generally see two trends: (i) The density profile follows $r^{-2}$ at inner radii but steepens at outer radii as they probe the time-evolving nature of $\dot{M}_w$ (inset), and (ii) The temperature at inner radii has a $T\propto r^{-3/4}$ dependence, expected for a steady $\rho\propto r^{-2}$ wind \citep{Piro20}.

Using the temperature profile, we obtain the color radius $r_c$, the outermost radius where the radiation and gas can be thermally coupled. We set the condition for thermal coupling by the condition on the thermalization optical depth \citep{Rybicki86},
\begin{eqnarray}
    \int_{r_c}^{r_{\rm out}}\sqrt{3\kappa_{\rm abs}(\kappa+\kappa_{\rm abs})}~\rho dr = 1.
\end{eqnarray}
For $\kappa_{\rm abs}$, we adopt a Kramer's opacity law motivated for bound-free/free-free absorption, with a sharp cutoff at low temperatures due to H/He recombination
\begin{eqnarray}
    \kappa_{\rm abs}(\rho, T) = 
    \begin{cases}
       C\rho T^{-3.5} & (T\geq T_c)\\
       C\rho T_c^{-3.5}(T/T_c)^\beta & (T<T_c),
    \end{cases}
\end{eqnarray}
where $C$, $T_c$ are dependent on the composition of the wind. Here we set $\beta=15$ to mimic the sharp cutoff on the opacity at low $T<T_c$ expected from recombination \citep[e.g.,][]{Faran19}.

When $r_c$ solved above has the relation $r_c>r_{\rm tr}$ (``thermalization-dominated temperature" in \citealt{Piro20}), the temperature of the emission $T_{\rm rad}$ would approximately be the wind temperature at $r_c$,
\begin{eqnarray}
    T_{\rm rad} \approx T(r=r_c)\ (r_c>r_{\rm tr}).
\end{eqnarray}
On the other hand if $r_c<r_{\rm tr}$ (``trapping-dominated temperature" in \citealt{Piro20}), $T_{\rm rad}$ is instead set by the temperature of the wind at $r_{\rm tr}$. For our typical cases in Figure \ref{fig:rho_T_profile}, we see from the right panel that the wind is in the thermalization-dominated regime, and $T_{\rm rad}$ being lower for larger $\dot{M}_w$ due to more efficient reprocessing.

The temperature information is complementary to the bolometric light curve, and is important for two reasons. First, bolometric light curves can be constructed only if sufficient multi-band spectral energy distribution (SED) of the precursor is available. While this is feasible in the era of LSST as long as the SED peak is near optical and slowly evolves over timescales of weeks, 
in many pre-LSST precursors such detailed SEDs are absent. The temperature estimates along with the bolometric luminosities enable constructing multi-band light curves, that can be more directly compared with observations. 

Second and more importantly, the temperature information helps break degeneracies in light curve modeling. The representative case is that of the mass transfer rate $\dot{M}_*$ and the accretion efficiency $\epsilon_{\rm acc}$ (or wind velocity $v_w=\epsilon_{\rm acc}^{1/2}c$). The energy output of the precursor traces $\epsilon_{\rm acc}\dot{M}_*$, so inference from bolometric light curve alone results in degeneracies between $\dot{M}_*$ and $\epsilon_{\rm acc}$. The emission temperature traces the density of the wind, and hence carries information of $\dot{M}_*/v_w\propto \epsilon_{\rm acc}^{-1/2}\dot{M}$. Thus incorporating both luminosity and temperature data helps constrain the mass loss history much better, as we demonstrate in Section \ref{sec:fitting_events} with actual events.

\subsection{Accounting for Dust}
We further include an approximate model of dust correction at optical wavelengths, that affects the observed SED by both extinction and reddening. The dust correction generally becomes important when the wind is dense \citep[large $\dot{M}_w/v_w$; e.g.,][]{Kochanek11}, and expands to a large radius where dust can condense. 

From the bolometric light curve $L_{\rm rad}$, we estimate the radius out to which dust sublimates \citep[e.g.,][]{Waxman00}
\begin{eqnarray}
    r_{\rm sub} &\approx& \left(\frac{L_{\rm rad}Q_{\rm abs/em}}{4\pi ac T_{\rm con}^4}\right)^{\!\!1/2} \nonumber \\
    &\sim& 6\times 10^{15}~{\rm cm}\nonumber \\
    &&\times\ Q_{\rm abs/em}^{1/2} \left(\frac{L_{\rm rad}}{10^{41}~{\rm erg\ s^{-1}}}\right)^{\!\!1/2}\left(\frac{T_{\rm con}}{10^3~{\rm K}}\right)^{\!\!-2},
\end{eqnarray}
where $T_{\rm con}$ is the dust condensation temperature, and $Q_{\rm abs/em}$ is the ratio of the absorption and emission coefficients. We then calculate the dust optical depth in V-band by
\begin{eqnarray}
    \tau_{\rm d,V} \approx 
    \begin{cases}
    \int_{r_{\rm sub}}^{r_{\rm out}} \kappa_{\rm d,V}\rho dr & (r_{\rm sub}< r_{\rm out}), \\
    0 & (r_{\rm sub}\geq r_{\rm out}).
    \end{cases}
\end{eqnarray}

\begin{figure}
    \centering
 \includegraphics[width=0.9\linewidth]{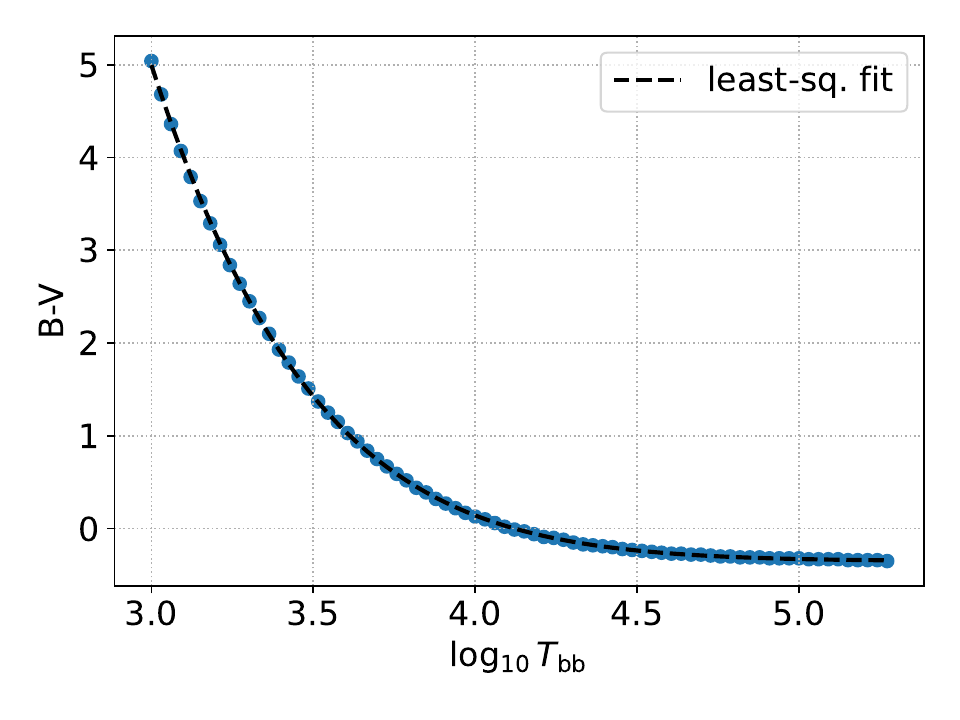}
    \caption{Relation between color $B-V$ and blackbody temperature $T_{\rm bb}$. Dashed line shows the fitting formula in equation (\ref{eq:B-V_from_T_bb})  used in our model for dust correction.}
    \label{fig:BminV_vs_Teff}
\end{figure}

From $\tau_{\rm d,V}$ we obtain the reddening $E(B-V)$ as
\begin{eqnarray}
    E(B-V) = \frac{A_V}{R_V} = \frac{1.086\tau_{\rm d,V}}{R_V},
\end{eqnarray}
where $A_V$ is the dust extinction (in mag) at V-band. To obtain the observed luminosity $L_{\rm obs}$ and temperature $T_{\rm obs}$, we first assume that the intrinsic SED (before dust extinction) is a Planck function with temperature $T_{\rm rad}$
\begin{eqnarray}
    L_{\lambda}\propto \frac{L_{\rm rad}}{T_{\rm rad}^4}\frac{\lambda^{-5}}{\exp[hc/\lambda k_B T_{\rm rad}]-1}.
    \label{eq:L_lambda}
\end{eqnarray}
Here $h,k_B$ are respectively the Planck and Boltzmann constants. We first obtain the observed color after dust extinction
\begin{eqnarray}
    (B-V)_{\rm obs} = (B-V)_{\rm rad} + E(B-V).
\end{eqnarray}
We map the color $B-V$ and blackbody temperature $T_{\rm bb}$ via conversion formulae obtained from non-linear least squares fit of tabulated data\footnote{The data used is a table of bolometric corrections, publicly available in the SuperNova Explosion Code (\citealt{Morozova15}; \url{https://stellarcollapse.org/index.php/SNEC.html})} (see Figure \ref{fig:BminV_vs_Teff}), 
\begin{eqnarray}
    B-V &=& a_0 + a_1\exp[a_2{\rm log}_{10}T_{\rm bb}+a_3({\rm log}_{10}T_{\rm bb})^2]\label{eq:B-V_from_T_bb}
\end{eqnarray}
where we determine the fitting parameters to be $(a_0, a_1, a_2, a_3)=(-0.3499, 89.7479, 0.1430, -0.3611)$. The inverse function of equation (\ref{eq:B-V_from_T_bb}),
\begin{eqnarray}
     &&\log_{10}T_{\rm bb} \nonumber \\
     &&=-\frac{a_2}{2a_3}\left[1+  \sqrt{1+\frac{4a_3}{a_2^2}\ln\left(\frac{B-V-a_0}{a_1}\right)}~\right] \label{eq:T_bb_from_B-V},
\end{eqnarray}
predicts $\log_{10}T_{\rm bb}$ within $2\%$ from the tabulated values for temperatures $3<\log_{10}T_{\rm bb}<5.2$ of our main interest. 

A hypothetical observer with simultaneous B and V-band data can derive an ``observed" temperature $T_{\rm obs}$ and bolometric luminosity $L_{\rm obs}$ by blackbody fitting. We calculate $T_{\rm obs}$ from the dust-corrected $(B-V)_{\rm obs}$ via equation (\ref{eq:T_bb_from_B-V}). We obtain $L_{\rm obs}$ by imposing that the V-band luminosity is reduced by a factor $\exp(-\tau_{\rm d,V})$ due to dust extinction. Using equation (\ref{eq:L_lambda}), this leads to
\begin{eqnarray}
    L_{\rm obs} &=& L_{\rm rad}\exp(-\tau_{\rm d, V}) \nonumber \\
    &&\times \left(\frac{T_{\rm obs}}{T_{\rm rad}}\right)^4 \frac{\exp[hc/\lambda_V k_B T_{\rm obs}]-1}{\exp[hc/\lambda_V k_B T_{\rm rad}]-1},
\end{eqnarray}
where we adopt $\lambda_V=5500\ {\rm \AA}$ for the characteristic wavelength of V-band. Our approach implicitly relies on the assumption that we observe these precursors mainly in the optical, which has been the case for existing events and will most likely be the case in the era of LSST.

\subsection{Summary of Model Parameters}
\label{sec:model_summary}
Table \ref{tab:parameters} summarizes the input parameters of our precursor model. We have eight parameters in the binary model ($\dot{M}_0, t_{\rm MT}, t_0, t_{\rm m}, \delta, r_{\rm in}, \epsilon_{\rm acc,0}, n_{\rm acc}$) that we aim to constrain, as well as composition-related parameters ($\kappa$, $C$, $T_c$) and dust-related parameters ($T_{\rm con}, Q_{\rm abs/em}, \kappa_{\rm d,V}$). 

For the composition-related parameters one could adopt characteristic values motivated from spectral information of the event, e.g. 
\begin{itemize}
    \item ($\kappa$, $C$, $T_c$) $=$ ($0.3\ {\rm cm^2\ g^{-1}}$, $4\times 10^{25}(Z/Z_\odot)$ cgs, $6000$ K) for H-rich events, and
    \item ($\kappa$, $C$, $T_c$) $=$ ($0.2\ {\rm cm^2\ g^{-1}}$, $2\times 10^{25}(Z/Z_\odot)$ cgs, $10000$ K) for H-poor, He-rich events.
\end{itemize}
As we consider thermalization between gas and radiation, we have adopted values of $C$ motivated from the Planck mean continuum opacity.\footnote{The value $C\approx 4\times 10^{25}$ cgs for solar abundance is found from radiative transfer modeling of \cite{Matsumoto21}. We expect the bound-free opacity to approximately scale with the metallcity $Z$ and the electron number density ($\propto (1+X_H)$, where $X_H$ is the hydrogen mass fraction).} In reality the thermalization radius is frequency-dependent, and by adopting the frequency-averaged opacity we are calculating the characteristic $r_c$ that governs where the SED peaks.

The dust parameters ($T_{\rm con}, Q_{\rm abs/em}, \kappa_{\rm d,V}$) depend on the grain size distribution and composition. For the subsequent analysis we adopt the following fiducial choices: $T_{\rm con}\approx 1200~$K, $Q_{\rm abs/em}\approx 10$, $R_V\approx 3.1$, and $\kappa_{\rm d,V}\approx 30(Z/Z_\odot)\ {\rm cm^2\ g^{-1}}$ that scales with metallicity $Z$, motivated by silicate dust with size distribution in the Galactic diffuse interstellar medium \citep[e.g.,][]{Lodders03,Draine11,Inoue20}. 

\begin{table}
    \centering
    \caption{Model parameters. We infer the former eight with our model, while the latter six are fixed for each event.}
    \begin{tabular}{cc}
    Variable & Description \\ \hline
    $\dot{M}_0$ & Mass transfer rate well before dynamical phase\\
    $t_{\rm MT}$ & Start time of mass transfer \\
    $t_0$ & Onset of the dynamical mass transfer phase \\
    $t_m$ & Time of binary merger \\
    $\delta$ & Mass transfer evolution power-law index  \\
    $r_{\rm in}$ & Wind launching radius \\
    $\epsilon_{\rm acc,0}$ & Initial accretion efficiency\\ 
    $n_{\rm acc}$ & Accretion efficiency evolution power-law index\\ \hline
    $\kappa$ & Total (scattering dominated) opacity  \\ 
    $C$ & Kramer's law scaling parameter \\
    $T_c$ & Recombination temperature \\ 
    $T_{\rm con}$ & Dust condensation temperature \\
    $Q_{\rm abs/em}$ & Ratio of dust absorption/emission coefficient\\
    $\kappa_{\rm d,V}$ & V-band dust opacity \\ \hline
    \end{tabular}
    \label{tab:parameters}
\end{table}

\section{Expected Ranges of Model Parameters}
\label{sec:applications}
So far the light curve model has been agnostic to the detailed binary systems or physics powering the precursor. In this section we aim to connect these two by forward modeling under a specific accretion model, to obtain rough insights for the model parameters in Section \ref{sec:model}. In particular, we examine values of $\epsilon_{\rm acc}$ and $r_{\rm in}$ that would most characterize the precursor emission.

\subsection{General framework for accretion energetics}
\label{sec:L_accretion_framework}
A framework we adopt here to calculate the accretion follows \cite{Tsuna24}, which coupled the mass transfer history to a one-zone model of \cite{Lu23} that calculates the fraction of mass forming the accretion disk versus that spills out from the binary via the L2 point. We define the latter fraction by $f_{\rm L2}$, with the former disk feeding rate being $\dot{M}_{\rm acc}=(1-f_{\rm L2})|\dot{M}_*|$. The one-zone model of \cite{Lu23} calculates $f_{\rm L2}$ as a function of binary separation $a_{\rm bin}$, mass transfer rate $|\dot{M}_*|$, and the binary mass ratio $q=M_\bullet/M_*$. 

As shown in \cite{Lu23}, a large $f_{\rm L2}\approx 1$ is expected when $|\dot{M}_*|\gtrsim 10^{-2}-10^{-3}(10^{-3}-10^{-4})~M_\odot\ {\rm yr}^{-1}$ for a $10~M_\odot$ ($1.4~M_\odot$) accretor, with variations due to abundance, accretor mass and separation. This threshold is mainly set by what most efficiently takes away energy from the disk, either radiation, inward advection, or outward advection through L2. As the latter becomes most efficient, $f_{\rm L2}$ will approach unity. For example, if one considers an H-poor abundance instead of solar abundance, this enhances the radiative cooling rate due to the reduced opacity and lowers $f_{\rm L2}$ (see their Appendix). A more massive accretor makes advection more efficient for a fixed disk radius, also reducing $f_{\rm L2}$. Overall, for high $|\dot{M}_*|$ we typically expect a disk feeding rate capped to some value, of $\dot{M}_{\rm acc}\sim (10^{-4}-10^{-2})~M_\odot\ {\rm yr^{-1}}$ depending on the binary system. This is still much larger than $\dot{M}_{\rm Edd}$ for stellar-mass compact objects, and as we discuss below we expect most of $\dot{M}_{\rm acc}$ to eventually escape without reaching the compact object. This justifies the approximation of $\beta\ll 1$ in equation (\ref{eq:Mdot_wind}).

Now we turn to modeling the accretion power using $\dot{M}_{\rm acc}$ obtained above. There are three key radii (from the compact object) that govern the energetics of the accretion. First is the outermost disk radii, set by circularization due to finite angular momentum of the L1 stream with respect to the accretor \citep{Lu23}
\begin{eqnarray}
    r_{\rm disk} &\approx& \frac{(1-x_{\rm L1})^4(1+q)}{q} a_{\rm bin}  \nonumber \\
    &\sim& (0.04{\rm -}0.1)a_{\rm bin}\ (0.01<q<0.8),
\end{eqnarray}
where $x_{\rm L1} \approx -0.0355(\log_{10}q)^2 - 0.251\log_{10}q + 0.500$ is the distance to L1 from the donor scaled by $a_{\rm bin}$, and corotation of the donor with the orbit is assumed. The second is the ``spherization radius", inside which the local accretion luminosity $\sim GM_\bullet\dot{M}_{\rm acc}/r$ exceeds the Eddington limit $4\pi GM_\bullet c/\kappa$ \citep{Shakura73,Begelman79},
\begin{eqnarray}
    r_{\rm sph} &\approx& \frac{\kappa \dot{M}_{\rm acc}}{4\pi c} \nonumber \\
    &\sim& 5\times 10^{10}\ {\rm cm}\left(\frac{\kappa}{0.3\ {\rm cm^2\ g^{-1}}}\right)\left(\frac{\dot{M}_{\rm acc}}{10^{-3}\ M_\odot {\rm yr}^{-1}}\right).
\end{eqnarray}
The last radii is the innermost radius of the accretion disk. We adopt the innermost stable circular orbit for a spinless compact object
\begin{eqnarray}
    r_{\rm ISCO} \approx \frac{6GM_\bullet}{c^2}\approx 12.5\ {\rm km}\left(\frac{M_\bullet}{1.4~M_\odot}\right),
\end{eqnarray}
which is also approximately the surface radius for a NS accretor \citep[e.g.,][]{Abbott18,Miller21}. 

We generally have the hierarchy $r_{\rm ISCO}\ll (r_{\rm sph}, r_{\rm disk})$. The inequality $r_{\rm sph}<r_{\rm disk}$ would mean that only the inner region with $r<r_{\rm sph}$ is radiatively inefficient and can potentially launch outflows, while $r_{\rm sph}>r_{\rm disk}$ means that the super-Eddington outflows could be launched everywhere in the disk.

For estimating the energetics from this accretion we make two simplifying prescriptions as done in related analytical modeling of super-Eddington accretion \citep[e.g.,][]{Fuller22,Tsuna25}. First, the accretion power is set by the (kinetic) energy of the outflows with a radially-dependent specific energy assumed to be the binding energy $\approx GM_\bullet/2r$. Second, the outflows reduce the inflow rate with radial dependence $\dot{M}(r)=\dot{M}_{\rm acc}(r/r_{\rm outflow})^{0.5}$, where $r_{\rm outflow}={\rm min}(r_{\rm sph}, r_{\rm disk})$ is the outermost radius where outflows develop. The power-law index of $0.5$ is motivated from recent large-scale simulations of radiatively inefficient accretion flows \citep{Cho24, Cho25, Guo24, Guo25}. Integration with $r$ leads to a total accretion power from the inner disk wind
\begin{eqnarray}
    L_{\rm in} &=& \int_{\rm r_{\rm ISCO}}^{r_{\rm outflow}} \left(-\frac{d\dot{M}(r)}{dr}\right) \frac{GM_\bullet}{2r}dr\nonumber\\
    &=& \frac{GM_\bullet \dot{M}_{\rm acc}}{2r_{\rm ISCO}}\left[\left(\frac{r_{\rm ISCO}}{r_{\rm outflow}}\right)^{\!\!1/2} - \left(\frac{r_{\rm ISCO}}{r_{\rm outflow}}\right)\right] \nonumber \\
    &\sim& 5\times 10^{40}\ {\rm erg\ s^{-1}}\left(\frac{\dot{M}_{\rm acc}}{10^{-3}\ M_\odot {\rm yr}^{-1}}\right)\left(\frac{r_{\rm outflow}}{10^4r_{\rm ISCO}}\right)^{\!\!-1/2},\label{eq:L_acc}
\end{eqnarray}
where we used $r_{\rm ISCO}\ll r_{\rm outflow}$ in the last equation. The inner disk wind of luminosity $L_{\rm in}$ and mass outflow rate $\dot{M}_{\rm acc}[1-(r_{\rm ISCO}/r_{\rm outflow})^{0.5}]\approx \dot{M}_{\rm acc}=(1-f_{\rm L2})|\dot{M}_*|$ collides with the outer, much slower L2 outflow at radius $\sim a_{\rm bin}$ and mass outflow rate $f_{\rm L2}|\dot{M}_*|$. 

We could simplify the outcome of the collision of these two outflows as a single, merged outflow with mass-loss rate $|\dot{M}_*|$. Under such a one-zone approximation, momentum and energy conservation give (e.g. Sec 2 of \citealt{Murase14})
\begin{eqnarray}
    \dot{M}_{\rm acc}\sqrt{\frac{2L_{\rm in}}{\dot{M}_{\rm acc}}} &\approx&  |\dot{M}_*|V_{\rm tot}, \\
    L_{\rm in} &\approx& \frac{|\dot{M}_*|V_{\rm tot}^2}{2}  + L_{\rm diss},
\end{eqnarray}
where $V_{\rm tot}$ is the velocity of the final merged outflow, and $L_{\rm diss}$ is the luminosity dissipated (converted to internal energy) which can be solved by eliminating $V_{\rm tot}$ as
\begin{eqnarray}
    L_{\rm diss} \approx L_{\rm in} - \frac{\dot{M}_{\rm acc}}{|\dot{M}_*|}L_{\rm in}\approx f_{\rm L2}L_{\rm in}.
\end{eqnarray}

Hence for $f_{\rm L2}\simeq 1$, we expect a fraction $\approx f_{\rm L2}$ of the inner disk wind's kinetic energy to be efficiently converted to internal energy by collision with the L2 outflow. The dissipation region will span a range of radii that is comparable to $a_{\rm bin}$. For $f_{\rm L2}\ll 1$, dissipation by the L2 outflow does not play a significant role in generating internal energy. We instead expect efficient dissipation to only occur around $r_{\rm outflow}$ where most of the mass is launched, by e.g. collision of the faster wind from the vicinity of NS/BH and the slower wind from around $r_{\rm outflow}$. So accretion power could still be dissipated efficiently, but at a much smaller radii of $r_{\rm outflow}$.

These together give us an informative guide for what to expect for the parameters $r_{\rm in}$ and $\epsilon_{\rm acc}$ important for our light curve model. Equating the dissipated luminosity with the budget of internal energy (per time) in our light curve model $\epsilon_{\rm acc}|\dot{M}_*|c^2/2$, we obtain
\begin{eqnarray}
    r_{\rm in} \sim  a_{\rm bin}, \  \epsilon_{\rm acc}\approx \frac{2f_{\rm L2} L_{\rm in}}{|\dot{M}_*|c^2} && (f_{\rm L2}\simeq 1) \label{eq:epsacc_fL21} \\
    r_{\rm in} \sim r_{\rm outflow},\ \epsilon_{\rm acc}\approx \frac{2L_{\rm in}}{|\dot{M}_*|c^2} && (f_{\rm L2}\ll 1) .\label{eq:epsacc_fL20}
\end{eqnarray}
This leads to a crude estimate of $\epsilon_{\rm acc}\sim 2L_{\rm in}/|\dot{M}_*|c^2$ across $f_{\rm L2}$, which we use in calculations in Section \ref{sec:applications_to_binaries}.

\begin{figure*}
    \centering
    \includegraphics[width=0.95\linewidth]{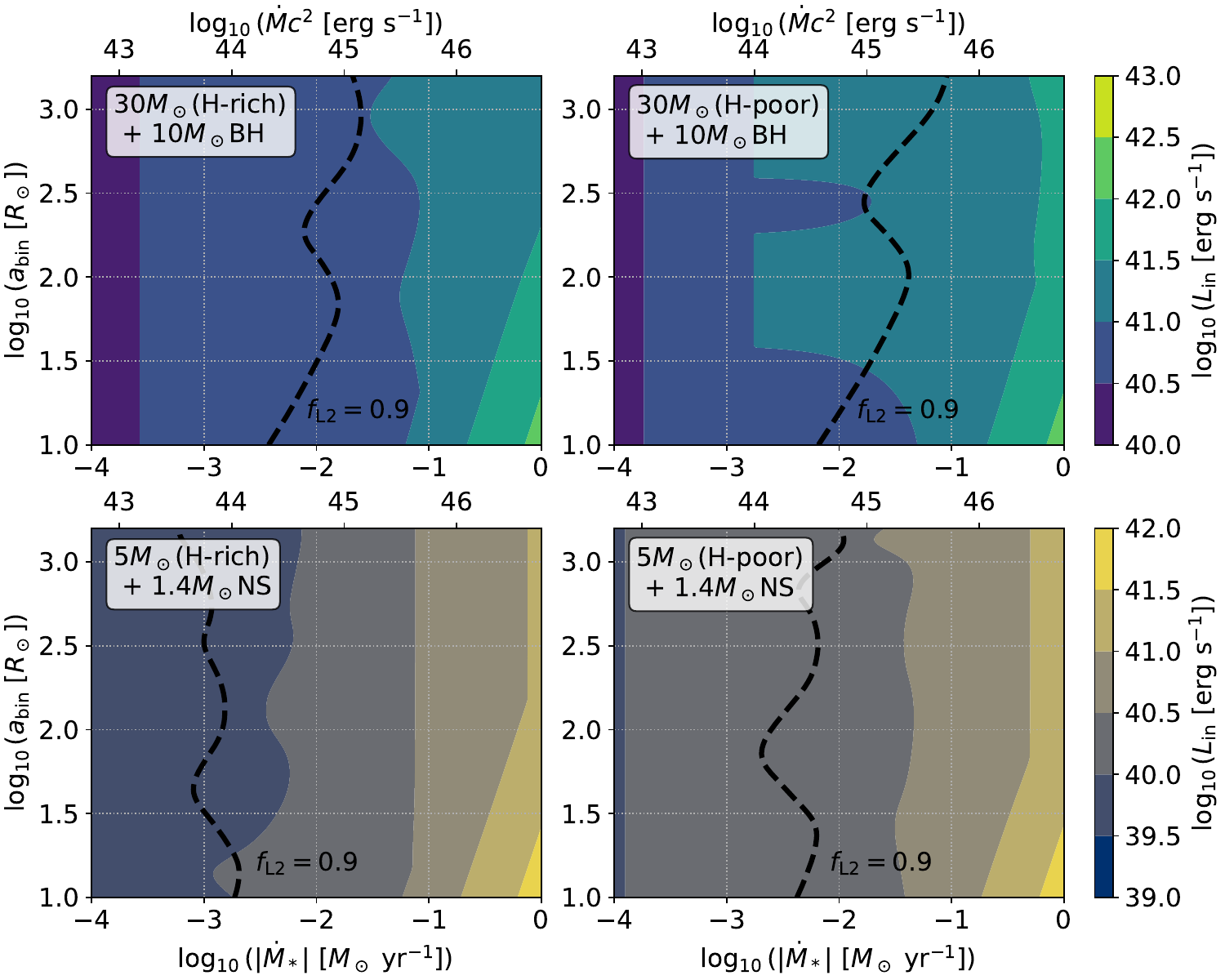}
    \caption{Inner wind (accretion) power $L_{\rm in}$ as a function of the mass-transfer rate $|\dot{M}_*|$ and binary separation $a_{\rm bin}$, under the model outlined in Section \ref{sec:applications}. We consider a NS (BH) accretor with mass $1.4~ (10)~M_\odot$, accreting from a $5~(30)~M_\odot$ donor with composition being H-rich ($X=0.7, Y=0.28, Z=0.02$) or H-poor ($X=0, Y=0.98, Z=0.02$). Overall, the model predicts $L_{\rm in}\approx 10^{40}$--$10^{42}$ erg s$^{-1}$ and $\epsilon_{\rm acc}\sim 2L_{\rm in}/|\dot{M}_*|c^2\sim 10^{-5}$--$10^{-3}$, with lower $\epsilon_{\rm acc}$ for higher $|\dot{M}_*|$ (see also Figure \ref{fig:Mdot_vs_eps}). The thick dashed lines show the contour of $f_{\rm L2}=0.9$ for $|\dot{M}_*|$, above which we expect most of the mass transferred through L1 to spill out through L2 without accreting.}
    \label{fig:L_wind}
\end{figure*}

\begin{figure}
    \centering  \includegraphics[width=\linewidth]{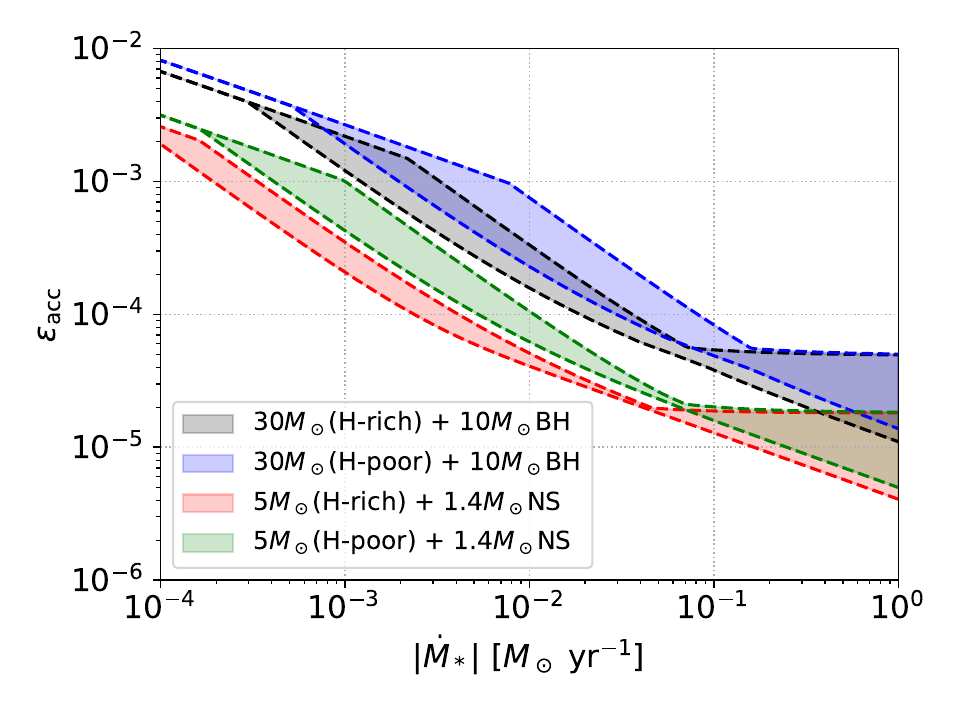}   
    \caption{Calculated accretion efficiency $\epsilon_{\rm acc}\approx 2L_{\rm in}/|\dot{M}_*c^2|$ as a function of $|\dot{M}_*|$, with a range given by variations of $a_{\rm bin}$ over the same range as in Figure \ref{fig:L_wind}. The efficiency is generally lower for less massive accretors, and higher $|\dot{M}_*|$ though with a more detailed evolution dependent on $a_{\rm bin}$.}
    \label{fig:Mdot_vs_eps}
\end{figure}

\subsection{Applications to Specific Binaries}
\label{sec:applications_to_binaries}
Having outlined the general framework, we turn to specific examples of binary systems. Here we choose fiducial sets of donor and accretor masses of $(M_*, M_\bullet)=(30~M_\odot, 10~M_\odot), (5~M_\odot, 1.4~M_\odot)$, representing a BH and NS accretor respectively in binaries with highly unequal mass ratios. We also consider the two extremes of solar abundance ($X=0.7, Y=0.28, Z=0.02$) and H-free composition ($X=0, Y=0.98, Z=0.02$), covering the expected range of massive star donors from hydrogen-rich stars (e.g. stars in Hertzsprung gap and supergiants, relevant to SN IIn) to helium stars (relevant to SN Ibn). This is admittedly not covering the broad landscape of possible binary properties, and we plan to expand this investigation in future study using realistic binary evolution models. Nevertheless the results are not sensitive to the mass ratio, compared to other key parameters like $|\dot{M}_*|$ and $M_\bullet$.

Figure \ref{fig:L_wind} shows the landscape of the inner wind power $L_{\rm in}$ as calculated in equation (\ref{eq:L_acc}). We see an overall trend of increasing $L_{\rm in}$ with $|\dot{M}_*|$, with a much weaker dependence on $a_{\rm bin}$ and wiggles reflecting temperature-dependent opacity features in the one-zone disk model. Comparing the four panels, we can also see variations with the compact object mass and the abundance, which directly reflect the dependence of $f_{\rm L2}$ on these as discussed in Section \ref{sec:L_accretion_framework}. Binaries with $10~M_\odot$ BH companions tend to have several times larger $L_{\rm in}$ than $1.4~M_\odot$ NS companions, due to the higher accretor mass allowing a near-proportionally larger $\dot{M}_{\rm acc}$.

Overall, the predicted accretion power in Figure \ref{fig:L_wind} spans a range of $L_{\rm in}\sim 10^{40}$--$10^{42}$ erg s$^{-1}$. When compared with $|\dot{M}_*|$ (or $|\dot{M}_*|c^2$ in the upper axis), this suggests an efficiency of $\epsilon_{\rm acc}\sim 2L_{\rm in}/|\dot{M}_*|c^2 \sim (10^{-5}$--$10^{-3})$ for these models. The efficiency being much below unity is due to a combination of (i) significant L2 mass loss leading to $\dot{M}_{\rm acc}<|\dot{M}_*|$, and (ii) most of the disk wind being launched from the outer part of the disk at speeds much less than $c$. 

We further show in Figure \ref{fig:Mdot_vs_eps} the range of $\epsilon_{\rm acc}$ for a given $|\dot{M}_*|$, with the former defined as $\epsilon_{\rm acc}\sim 2L_{\rm in}/|\dot{M}_*|c^2$. The efficiency is generally a declining function of $|\dot{M}_*|$ albeit with a non-trivial evolution, reflecting the non-trivial dependence of $f_{\rm L2}(|\dot{M}_*|)$ as well as $r_{\rm outflow}(\dot{M}_{\rm acc})$. 

Despite the apparently low efficiency, the accretion power is sufficient to reproduce the range of luminosities of the observed precursors of interacting SNe, as we demonstrate for some events in Section \ref{sec:fitting_events}. While we do not claim that all the observed precursors are from binary mergers, this suggests that future observations by e.g. LSST would help unveil the landscape of such events and probe the cataclysmic endpoints of massive star binaries.

\subsection{Potential Uncertainties}
\label{sec:caveats}
In the above, we roughly estimated the energetics of the accretion and the radiative efficiency of the outflow under a one-zone framework, tailored to our light curve model in Section \ref{sec:model} constructed under spherical symmetry. The details on dissipation of the inner disk wind by the L2 outflow can depend on the specific geometry of these two components. \cite{Pejcha16b} discusses that the L2 outflow's geometry is set by the photon diffusion timescale in the outflow ($t_{\rm diff,L2}\sim 3\kappa f_{\rm L2}|\dot{M}_*|/4\pi v_{\rm out, L2} c)$ and the outflow's expansion timescale ($t_{\rm exp,L2}\sim a_{\rm bin}/v_{\rm out, L2})$. The ratio of these two timescales is
\begin{eqnarray}
\frac{t_{\rm diff,L2}}{t_{\rm exp,L2}} &\approx & \frac{3\kappa f_{\rm L2}|\dot{M}_*|}{4\pi a_{\rm bin}c} \nonumber \\
&\sim& 2\left(\frac{\kappa}{0.3\ {\rm cm^2\ g^{-1}}}\right)\!\left(\frac{f_{\rm L2}|\dot{M}_*|}{0.1~M_\odot\ {\rm yr}^{-1}}\right)\left(\frac{a_{\rm bin}}{100~R_\odot}\right)^{-1} .
\label{eq:ratio_tdiff_texp}
\end{eqnarray}
If $t_{\rm diff,L2}/t_{\rm exp,L2}\gtrsim 1$, we expect the L2 outflow to be quasi-spherical. The dissipation of the disk wind energy is efficient soon after interacting with the L2 outflow, i.e. $r_{\rm in}\sim a_{\rm bin}$ as we have assumed above. If $t_{\rm diff,L2}\lesssim t_{\rm exp,L2}$ and radiative cooling is fast, we expect that the geometry of the L2 outflow is set by a more complicated interplay of radiative cooling and irradiation by the compact accretor (as well as the donor). If the L2 outflow is equatorial, the light curve could be powered by two components, from the disk wind interacting with the L2 outflow at equatorial angles and from the disk wind itself at polar angles away from the L2 material.

Our estimate for the accretion power is conservative in the sense that we do not include the contribution from material that actually reaches the inner edge of the disk. For example this could be a jet launched by a spinning BH \citep{Blandford77}. While we believe an extremely small natal spin predicted for the first-born BH \citep{Fuller19} makes this sub-dominant for most cases\footnote{For example, an estimate of the jet luminosity using a (rather optimistic MAD) model of \cite{Lowell24} results in $L_{\rm jet}\sim 4\times 10^{-3}\dot{M}(r=r_{\rm ISCO})c^2$ for a BH spin of $0.01$ \citep{Fuller19}. This is $\gtrsim 10$ times smaller than the estimate from equation (\ref{eq:L_acc}) of $L_{\rm in}\approx (1/12)\dot{M}(r=r_{\rm ISCO})c^2$ for $r_{\rm ISCO}=6GM_\bullet/c^2 \ll r_{\rm outflow}$.}, even a moderate BH spin may help explain the most extreme precursors requiring luminosities of $L_{\rm acc}\gtrsim 10^{42}$ erg s$^{-1}$. Another potential source is surface accretion in case of strongly magnetized NSs, which may be as high as $\sim 10^{39}$--$10^{40}$ erg s$^{-1}$ \citep[e.g.,][]{Mushtukov15} and could make a significant contribution if it is tapped to the disk wind. However it is uncertain if strong $B$-fields are realized for the NSs of our interest, which likely undergo strong accretion ($\dot{M}_{\rm NS}\gg \dot{M}_{\rm Edd}$) and magnetic field burial over their evolution \citep[e.g.,][]{Konar17}.

This framework has a key assumption of steady-state, with the evolution of $|\dot{M}_*|$ being the slowest process. This is valid for most of the mass transfer phase, as the accretion processes at the inner disk is expected to be fast. However, the last runaway phase of dynamical mass transfer significantly changes $|\dot{M}_*|$ over the orbital period, and the hydrodynamics governing L2 mass loss becomes a poor approximation due to loss of co-rotation. In our framework ${\dot{M}_*}/(d\dot{M}_*/dt) \approx (t_m-t)/\delta$ at the dynamical phase (equation \ref{eq:Mdot_formula}), and hence our light curve parameters become more uncertain when the time till merger $t_m-t$ is comparable to the orbital period.

\section{Applications to Observed Long-rising Precursors of Interacting Supernovae}
\label{sec:fitting_events} 
As a demonstration of our our light curve model in Section \ref{sec:model}, we compare the model to the light curves of observed long-rising precursor events. In this work we consider three events, SN 2023zkd, SN 2023fyq, and SN 2021qqp, which displayed year(s)-long precursor emission followed by a ramp-up to the optical peak. Such events are promising candidates of binary mergers, in which our model can be applied to the precursor emission.

For model fitting we use the Markov Chain Monte Carlo sampling Python package \texttt{emcee} \citep{Foreman-Mackey13}. We apply broad uniform priors for the model parameters $\{\log \dot{M}_0, t_{\rm MT}, t_0, t_m, \delta, \log r_{\rm in}, \log \epsilon_{\rm acc,0}, n_{\rm acc}\}$, with ranges listed in Table \ref{tab:fitting_results}. We first use the maximum likelihood method to determine the initial guess of these parameters. We then initialize 64 walkers at random positions around the initial guess with relative noise of $10^{-2}$. The sampling is done for 5000 steps per walker, with a burn-in period of 2000 steps. 

After a detailed discussion of the inferred parameters of the three events in Section \ref{sec:2023zkd}--\ref{sec:2021qqp}, in Section \ref{sec:progenitor_constrints} we consider the probable progenitors of each event based on our merger scenario and previous studies. Finally, in Section \ref{sec:double_peaks} we discuss the origin of the peculiar double-peaked light curve morphology, which is rare in interacting SNe but interestingly seen in all three events.

\begin{table*}
    \centering
    \caption{Priors of the model parameters, and posteriors from our MCMC fitting of three events. For definition of each parameter see Table \ref{tab:parameters}. Detailed corner plots are shown in the Appendix.}
    \renewcommand{\arraystretch}{1.6}
    \begin{tabular}{cc|ccc}
    Variable & Prior & SN 2023zkd & SN 2023fyq & SN 2021qqp \\ \hline
    $\log_{10}[\dot{M}_0/(M_\odot{\rm yr}^{-1})]$ & [-7,1] & $-1.29^{+0.60}_{-0.56}$ & $-1.57^{+0.05}_{-0.05}$ & $-1.59^{+0.25}_{-0.24}$ \\
    $t_{\rm MT}$ &  [-$10^4$ days, first detection]& $-19.28^{+7.58}_{-5.76}$  yr & $-16.03^{+7.27}_{-7.62}$ yr  & $-17.47^{+7.12}_{-6.93}$ yr \\
    $t_0$ &  [$t_{\rm MT}$, 0]& $-6.36^{+4.43}_{-8.42}$ yr & $-0.17^{+0.04}_{-0.05}$ yr  & $-4.11^{+2.12}_{-3.84}$ yr \\
    $t_m$ &  [$t_0, 0$] & $-51.79^{+35.27}_{-41.49}$ day & $-7.75^{+3.35}_{-2.42}$ day & $-18.19^{+12.86}_{-19.74}$ day\\
    $\delta$ & [13/11,2] & $1.50^{+0.32}_{-0.21}$ & $1.51^{+0.30}_{-0.23}$ & $1.45^{+0.30}_{-0.19}$ \\
    $\log_{10} (r_{\rm in}/{\rm cm})$ & [6,15] & $12.67^{+0.86}_{-1.28}$ & $12.86^{+0.41}_{-0.55}$ & $12.69^{+0.62}_{-0.95}$ \\
    $\log_{10}\epsilon_{\rm acc,0}$ & [-8,0] & $-3.68^{+0.51}_{-0.17}$ & $-4.07^{+0.03}_{-0.03}$ &  $-3.75^{+0.31}_{-0.20}$\\
    $n_{\rm acc}$ & [-1,0] & $-0.18^{+0.12}_{-0.16}$ & $-0.30^{+0.09}_{-0.09}$ & $-0.28^{+0.11}_{-0.13}$ \\ \hline \hline
     \multicolumn{2}{c|}{Light curve data} & \cite{Gagliano25b} & \cite{Dong24} & \cite{Hiramatsu24} \\
    \multicolumn{2}{c|}{$t=0$ (MJD)}  & 60290.6 & 60143.77 & 59438.33 
    \end{tabular}
    \vspace{2mm}\\
    {Note: Epochs $t_{\rm MT}, t_0, t_m$ are relative to the zero point ($t=0$) defined in the final row.}
    \label{tab:fitting_results}
\end{table*}

\subsection{Type IIn SN 2023zkd}
\label{sec:2023zkd}

\begin{figure*}
    \centering
    \includegraphics[width=0.96\linewidth]{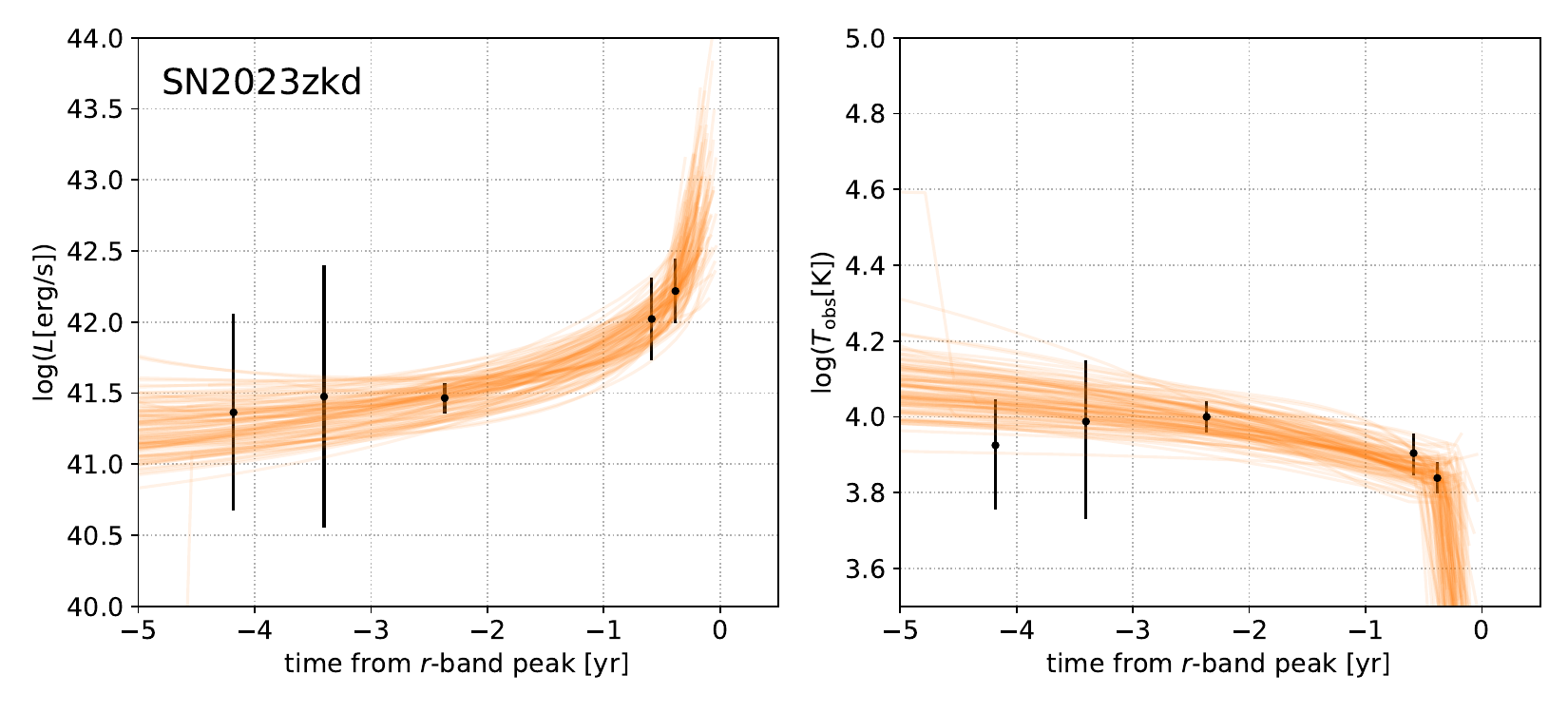}
    \caption{Reconstructed light curves and temperature evolution from 100 random posterior draws, for the case of SN 2023zkd. The five data points from the blackbody fit of \cite{Gagliano25b} are used for the MCMC fit.}
    \label{fig:2023zkd_LCfit}
\end{figure*}

\begin{figure*}
\centering
    \begin{minipage}{.49\textwidth}
    \centering
    \includegraphics[width=\linewidth]{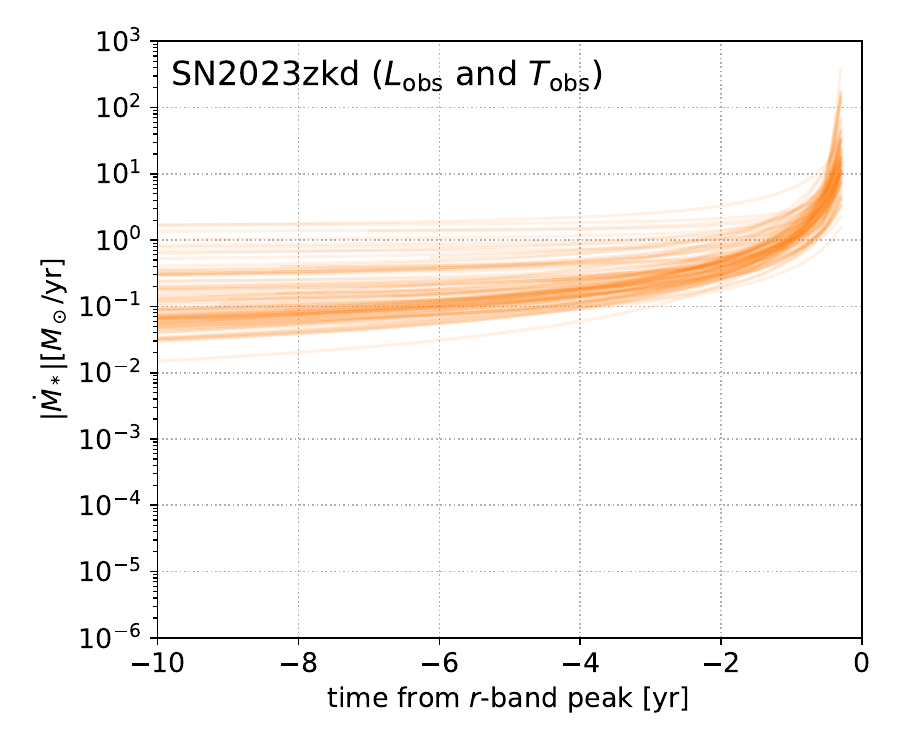} 
    \end{minipage} 
    \begin{minipage}{.49\textwidth}
    \centering
    \includegraphics[width=\linewidth]{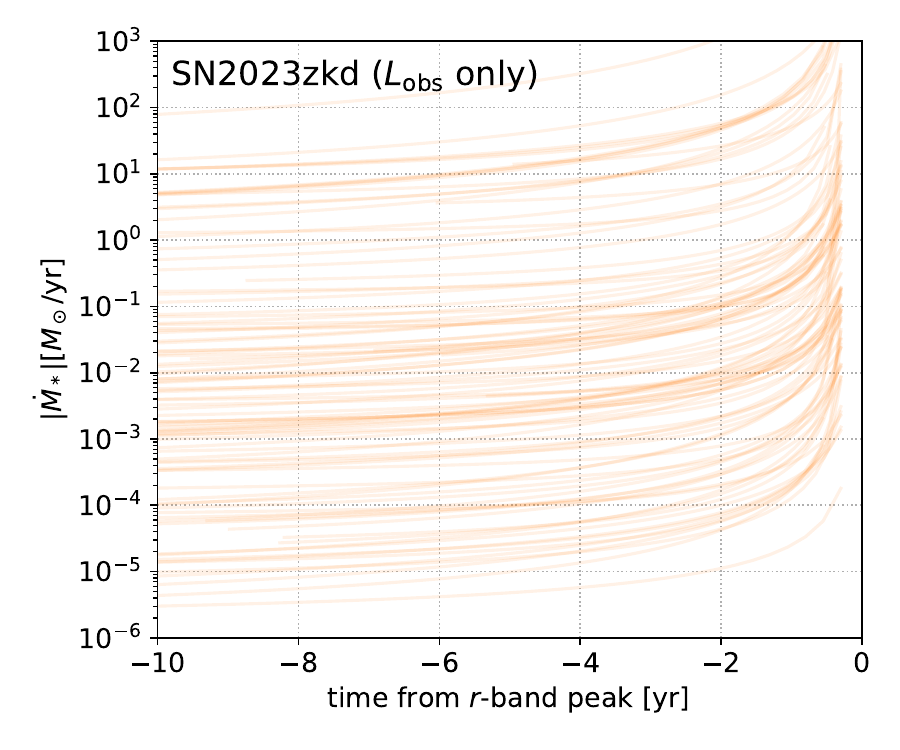} 
    \end{minipage} 
    \caption{Reconstructed mass transfer history for SN 2023zkd, showing two cases where both luminosity and temperature evolution were used for fitting as done in Figure \ref{fig:2023zkd_LCfit} (left), and where only the luminosity was used (right). Using both luminosity and temperature removes the degeneracy between $\dot{M}_*$ and $\epsilon_{\rm acc}$, and constrains the mass transfer history much better.}
    \label{fig:2023zkd_Mdot}
\end{figure*}

SN 2023zkd \citep{Gagliano25b} is a Type IIn SN that exhibited steady, luminous ($\approx -15$ mag) precursor emission from 4 years before the optical peak. The light curve displayed two peaks, separated by $\sim 200$ days in $r$-band. The CSM inferred from the SN phase is massive (a few to several $~M_\odot$), and multi-phase with varying hydrogen/helium-rich composition. From the properties of the CSM and the long-rising luminous precursor, \cite{Gagliano25b} concluded this event to be a merger between a partially stripped massive star and a BH companion.

We specifically fit their blackbody estimates $(L_{\rm obs}, T_{\rm obs})$ at the precursor phase (see their Figure 13), from day $-1527$ to day $-140$ with respect to the first $r$-band peak. For the model, we adopt the opacities for the H-rich composition (Section \ref{sec:model_summary}). By modeling the host galaxy emission, \cite{Gagliano25b} found a stellar metallicity of $\log_{10} (Z_*/Z_\odot)=-1.18^{+0.03}_{-0.01}$ and a gas metallcity of $\log_{10}(Z_{\rm gas}/Z_\odot)=-0.50^{+0.05}_{-0.05}$. We adopt $Z=0.3Z_\odot$ from the gas metallicity, as this more directly traces recent star formation.

Figure \ref{fig:2023zkd_LCfit} shows light curve fits by 100 random draws from the posterior distribution, with their inferred values shown in Table \ref{tab:fitting_results} (for the corner plot see Figure \ref{fig:2023zkd_corner}). The initial efficiency $\epsilon_{\rm acc,0}$ is constrained to $\sim 2\times 10^{-4}$, with a weak but significant evolution over time. We typically find a wind velocity of $\sqrt{\epsilon_{\rm acc}}c\sim 1000$--$4000$ km s$^{-1}$ around the end of the precursor phase at $\lesssim 2$ yr before merger, when the CSM seen in spectra may have originated based on the light curve fits of \cite{Gagliano25b}. While the error is large, this is broadly consistent with the CSM velocity measured from the spectral line widths in the SN phase (1200-2200 km s$^{-1}$).

Figure \ref{fig:2023zkd_Mdot} shows the reconstructed mass transfer history. The left panel shows the case where both $L_{\rm obs}$ and blackbody temperature $T_{\rm obs}$ were used for fitting, and the right panel shows a hypothetical case where only the bolometric light curve $L_{\rm obs}$ was used. As discussed in Section \ref{sec:model_temp}, the latter case results in a degeneracy between $|\dot{M}_*|$ and $\epsilon_{\rm acc}$, and as a result the mass transfer history $|\dot{M}_*|$ is poorly constrained. The case with temperature information helps break this degeneracy, and demonstrates the importance of temperature constraints from multi-band light curves.

The fitting constrains the mass transfer history to $|\dot{M}_*|\sim 0.1$-- a few $M_\odot$ yr$^{-1}$ over the precursor phase of $0.5$--$4$ years before the merger.  The distribution of time-integrated mass lost from the first to last data point peaks at $\sim 3~M_\odot$, also comparable to the CSM mass from light curve modeling in the SN phase done in \cite{Gagliano25b}. The information of CSM mass and velocity inferred from the SN phase were not imposed in our model that instead only fits the precursor phase. This supports the interpretation of \cite{Gagliano25b} that the precursor was powered by a nearly-merging binary with a compact object accretor.

We note that the required $\epsilon_{\rm acc}\sim 10^{-4}$ for $|\dot{M}_*|\sim 0.1-1~M_\odot\ {\rm yr}^{-1}$ during the precursor phase is a few to several times larger than the estimations in Section \ref{sec:applications} for a $10~M_\odot$ BH accretor, and even more for a NS accretor (Figure \ref{fig:Mdot_vs_eps}). This may indicate a BH heavier than considered in Figure \ref{fig:Mdot_vs_eps}, which may be motivated by the low metallicity of the system \citep[e.g.][]{Belczynski10,Andrews25}. Another possibility is an additional energy component not considered in the accretion model of Section \ref{sec:applications}, such as jets from a spinning BH. 

\subsection{Type Ibn SN 2023fyq}
\begin{figure*}
    \centering
    \includegraphics[width=0.96\linewidth]{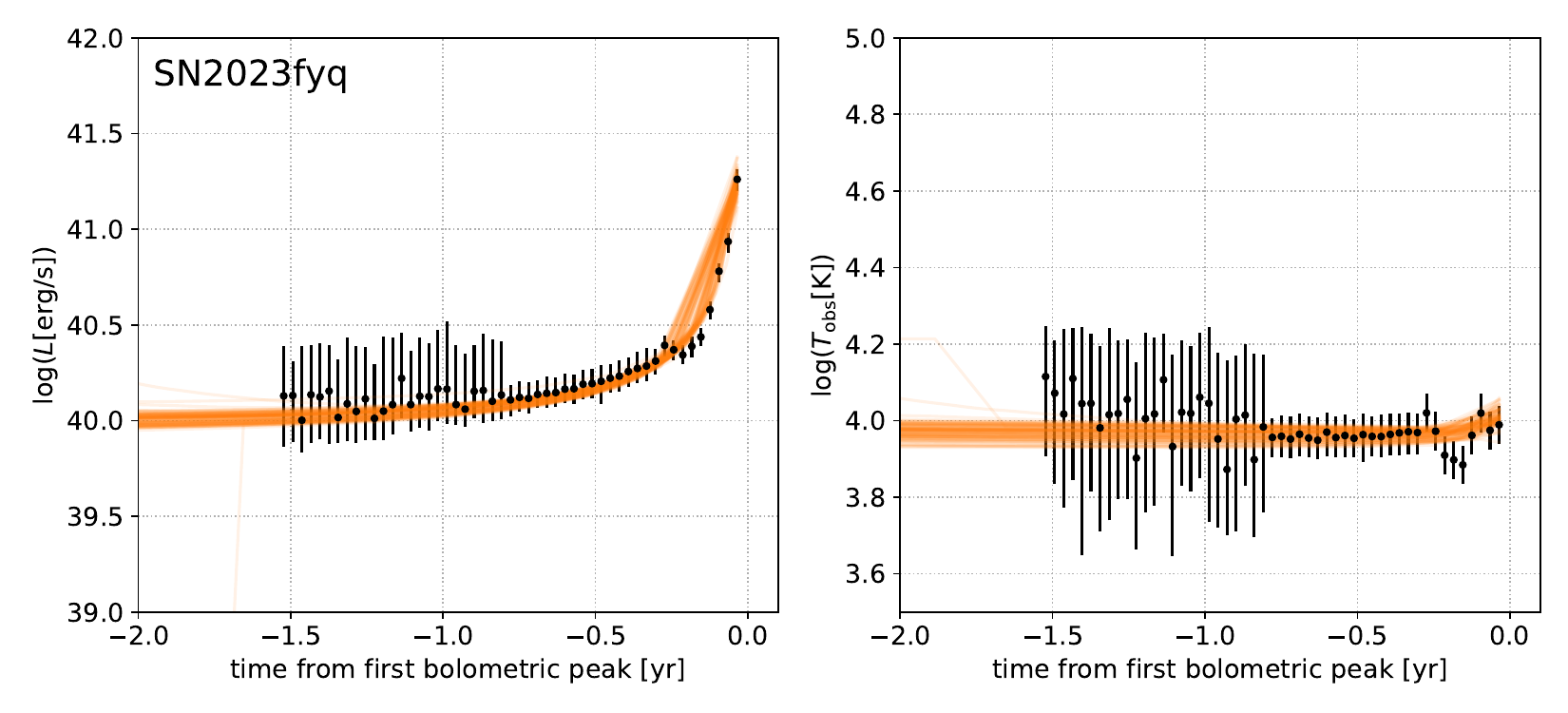}
    \caption{Same as Figure \ref{fig:2023zkd_LCfit}, but a light curve fit for SN 2023fyq. The data points are from the blackbody fit of \cite{Dong24}.}
    \label{fig:2023fyq_LCfit}
\end{figure*}
\begin{figure*}
\centering
    \begin{minipage}{.49\textwidth}
    \centering
    \includegraphics[width=\linewidth]{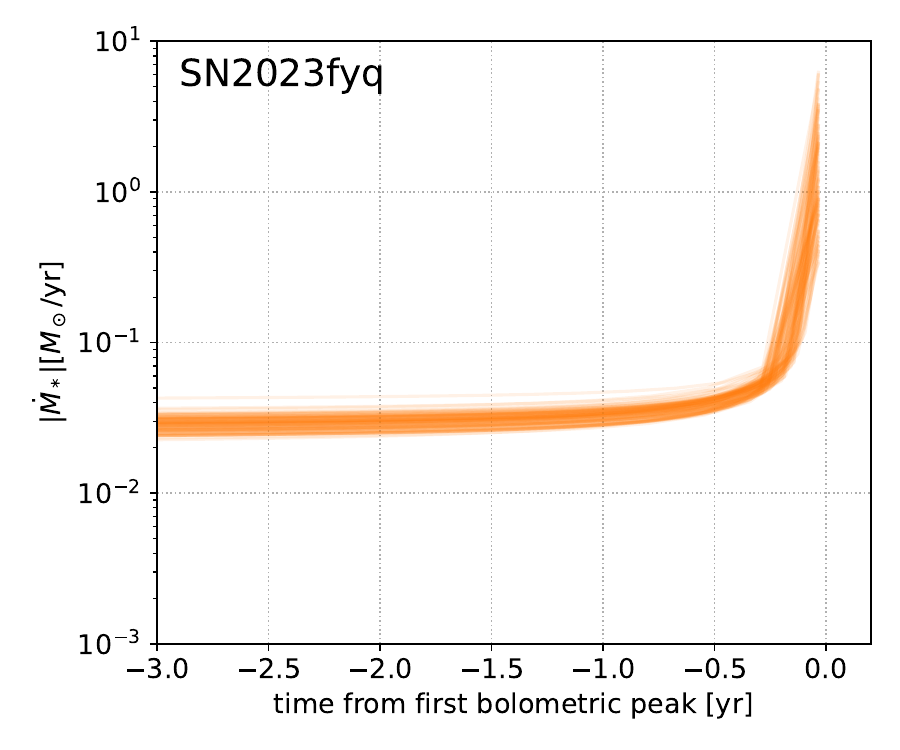} 
    \end{minipage} 
    \begin{minipage}{.49\textwidth}
    \centering
    \includegraphics[width=\linewidth]{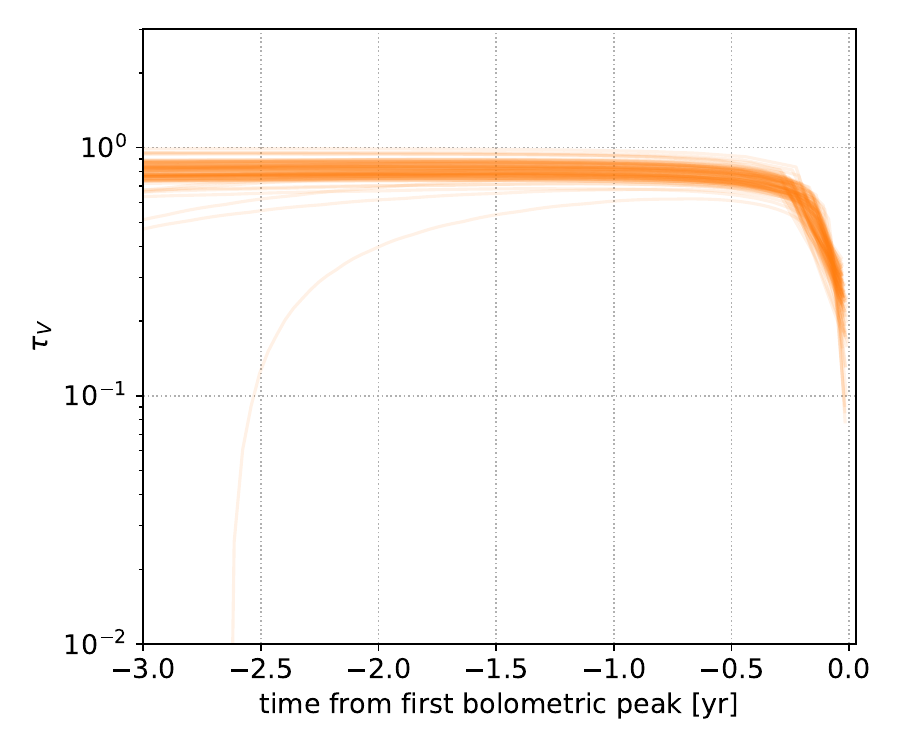} 
    \end{minipage} 
    \caption{The reconstructed mass-transfer rates and evolution of the $V$-band optical depth for SN 2023fyq. The sharp rise in luminosity near merger leads to dust sublimation and a drop in the dust optical depth.}
    \label{fig:2023fyq_Mdot_tau}
\end{figure*}
SN 2023fyq \citep{Brennan24,Dong24} is a Type Ibn SN where long-rising precursor emission was detected from 3 -- 5 years before the final explosion. Optical spectra at $\lesssim 50$ days before peak show P-Cygni features in helium lines, hinting presence of dense CSM with velocity of $\sim 1700$--$2500$ km s$^{-1}$ likely ejected near the end of this precursor phase \citep{Brennan24}. This SN is the first SN Ibn detected in the radio \citep{Baer-Way25}, with late-time observations confirming strong mass loss within the last 5--10 years from the SN. 

Similar to SN 2023zkd, we fit the blackbody estimates ($L_{\rm obs}, T_{\rm obs})$ from \cite{Dong24}\footnote{We have put a 5\% floor on the error of the temperature estimates, based on comparisons with the temperature estimates of \cite{Brennan24}.} in the precursor phase, defined here by $>11$ days before $r$-band maximum at MJD 60143.8. We adopt the opacities for the H-poor composition (Section \ref{sec:model_summary}), with a metallicity $Z\approx 0.6Z_\odot$ based on gas metallicity measurements of the star-forming regions of the host \citep{Hong26}. In addition to the priors in Table \ref{tab:fitting_results}, we impose a constraint on the CSM velocity $v_w$ from pre-explosion spectra of \cite{Brennan24}. We constrain $v_w$ as $1700<v_w/[{\rm km\ s^{-1}}]<2500$ at $40$ days before MJD 60155.1 (zero-point in \citealt{Brennan24}), when the P-Cygni features are clearly seen with no significant evolution before and after.

Figure \ref{fig:2023fyq_LCfit} shows the reconstructed light curves from 100 random posterior draws, and Figure \ref{fig:2023fyq_corner} shows the corresponding corner plots. Compared to SN 2023zkd, the larger number of data points and the constraint on pre-SN velocity lead to much tighter constraints on the key model parameters. For example, the initial mass-transfer rate and the accretion efficiency are both constrained at the $\sim 10\%$ level, to be $\dot{M}_0\approx 2.7\times 10^{-2}~M_\odot\ {\rm yr}^{-1}$ and $\epsilon_{\rm acc,0}\approx 8.5\times 10^{-5}$ (or initial velocity $v_w\approx 2600\ {\rm km\ s^{-1}}$, which drops to what is observed in the pre-explosion spectra). This shows the importance of spectroscopy {\it in the precursor phase}, which we believe could be routinely possible in the LSST era when we can find some precursors in real-time.

The left panel of Figure \ref{fig:2023fyq_Mdot_tau} shows the reconstructed mass-transfer history. The evolution is broadly similar to the earlier merger model of \cite{Tsuna24}, where they considered a low-mass helium star donor and a NS accretor. As discussed in \cite{Baer-Way25}, the merger scenario has a strength of simultaneously explaining the largely discrepant mass-loss rates inferred in early optical ($\sim 1~M_\odot$ yr$^{-1}$ at weeks before SN) and late-time radio ($\sim 10^{-2}~M_\odot$ yr$^{-1}$ at years before SN). 

We note that our inferred CSM density $(\propto \dot{M}_w/v_w)$ from mass loss years before the SN is a factor of 3--4 larger than the radio-based estimate of \cite{Baer-Way25}. The radio-based mass-loss estimates depend on both the CSM composition and the uncertain shock velocity ($\propto v_{\rm sh}^{1.5}$, for free-free dominated absorption). \cite{Baer-Way25} mention that the latter is adopted from the late-time spectral line width, and may be an underestimate. Perhaps more importantly, the observed radio signal is attenuated by free-free absorption, and mass-loss estimates can be sensitive to asymmetry in the CSM. The radio emission and reprocessed optical emission respectively probe low-density regions and high-density regions, and may explain the discrepancy. Asymmetric CSM could be due to a mild asymmetry in the L2 outflow (see scaling in equation \ref{eq:ratio_tdiff_texp}), and is also suggested from the multi-component features in the optical spectra \citep{Brennan24,Dong24}. 

The right panel of Figure \ref{fig:2023fyq_Mdot_tau} shows the V-band dust optical depth based on our dust modeling. The model predicts some dust reddening during the precursor phase, though the sharply rising luminosity leads to significant dust sublimation and reduction in the optical depth. There has been a detection of infrared emission from serendipitous JWST imaging at about a month before the bolometric peak \citep{Taggart24}, which may be due to reprocessing by the surviving dust. Our inferred $\tau_V\sim 0.2$ at around this JWST epoch leads to an order of magnitude estimate of the reprocessed infrared luminosity of $\sim 10^{40}$ erg s$^{-1}$, although more detailed treatment of dust physics and radiative transfer is required for accurate estimates.

\subsection{Multi-band fitting of Type IIn SN 2021qqp}
\label{sec:2021qqp}

\begin{figure*}
    \centering
    \includegraphics[width=0.96\linewidth]{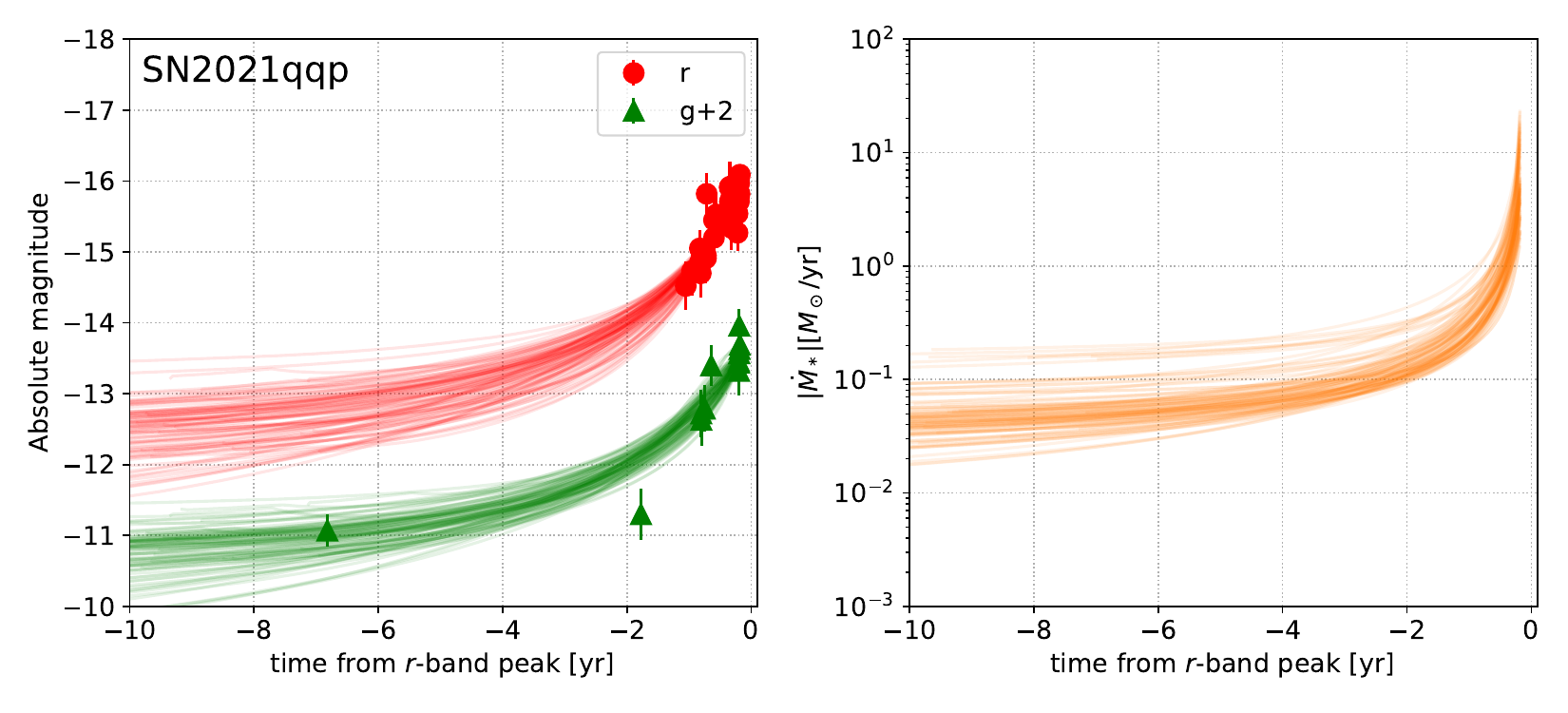}
    \caption{Multi-band light curves fits and reconstructed mass-transfer histories for SN 2021qqp. The data points in the left panel are ZTF data compiled by \cite{Hiramatsu24}, up to $\approx 65$ days before the optical peak.}
    \label{fig:2021qqp_LCfit}
\end{figure*}

SN 2021qqp \citep{Hiramatsu24} is a Type IIn SN that showed a steady-rising precursor evolving from $\approx -14$ mag to $-16$ mag in $r$-band over $300$ days. Like SN 2023zkd, the light curve displays two peaks that are separated by $\sim 300$ days in $r$-band. The SN phase displays interaction with a multi-phase CSM with different velocities, $\approx 1300$ km s$^{-1}$ at seen at first peak and $\approx 2500$ km s$^{-1}$ at second peak.

As the blackbody fits are not available during the long-rising precursor phase, we instead develop a multi-band fit to the precursor using our luminosity and temperature estimates. We extract the $g$- and $r$-band light curves from the Zwicky Transient Facility (ZTF), up to 65 days before the first $r$-band peak. \cite{Hiramatsu24} defines this last epoch as the end of the precursor phase, based on the sharp transition of the light curve morphology, and spectra taken soon after (35 days before peak) that display multi-component hydrogen features commonly seen in SN IIn. 

From our model, we calculate the AB magnitudes in $g$- and $r$-bands by
\begin{eqnarray}
    M_{g,r} = -2.5\log_{10}\left[\frac{L_{\nu=\nu_g,\nu_r}/4\pi(10\ {\rm pc})^2}{3631\ {\rm Jy}}\right],
\end{eqnarray}
where we adopt $(\nu_g, \nu_r)$ from the effective frequencies of the filters\footnote{From \url{http://svo2.cab.inta-csic.es/svo/theory/fps/}}, and calculate $L_\nu$ assuming a greybody with total luminosity $L_{\rm obs}$ and temperature $T_{\rm obs}$
\begin{eqnarray}
    L_{\nu} \approx \frac{L_{\rm obs}}{\sigma_{\rm SB} T_{\rm obs}^4} \frac{2\pi h \nu^3}{c^2}\frac{1}{ \exp[h\nu/k_BT_{\rm obs}]-1},
\end{eqnarray}
where $\sigma_{\rm SB}$ is the Stefan-Boltzmann constant.
We adopt the H-rich composition (Section \ref{sec:model_summary}), with a metallicity of $Z\approx 0.5Z_\odot$ based on fitting of the host galaxy photometry by \cite{Hiramatsu24} ($\log_{10}(Z/Z_\odot) = -0.29^{+0.28}_{-0.34}$).

Figure \ref{fig:2021qqp_LCfit} shows the multi-band fitting and the reconstructed mass transfer history. Overall we find very similar parameter constraints as SN 2023zkd, except the evolution in $|\dot{M}_*|$ being lower by a factor of a few. The reconstructed light curves do not fit the $g$-band data well at $\approx -1.8$ years, likely due to the fitting being heavily weighted towards the ramp-up part with many more data points. The CSM velocity in our model decreases as the binary approaches merger, with e.g. a $1\sigma$ range of $v_w\approx1800$--$5900$ km s$^{-1}$ at two years before merger and $v_w\approx 1100$--$3600$ km s$^{-1}$ at the last data point. These two are consistent with the observed CSM velocity estimates (though with large uncertainties in the models), and may explain the observed trend of increasing CSM velocity later in the SN phase as the shock sweeps through material emitted earlier.

\subsection{Implications for the binary progenitors of the observed precursor events}
\label{sec:progenitor_constrints}
The inferred mass transfer history should carry information about the binary, as described in equation (\ref{eq:Mdot_estimate}). However degeneracies remain in the donor's mass, radii and structure as well as the mass ratio, and decoding these requires predictions of the parameter space  from grids of binary evolution models. While understanding this is deferred to future work, we nevertheless attempt to make qualitative discussions on the progenitors, and assess the proposed scenarios for individual events.

For the H-rich SN 2023zkd and SN 2021qqp, the inferred mass-loss rate during the dynamical phase is large, e.g. $\sim 0.3$--$1~M_\odot$ yr$^{-1}$ at 1 year before merger (Figure \ref{fig:2023zkd_Mdot} and \ref{fig:2021qqp_LCfit}). From equation (\ref{eq:Mdot_estimate}), this requires a fairly evolved massive star donor, with large radii of $\gtrsim 10^2~R_\odot$ and mass of 10s of $M_\odot$, likely in the Hertzsprung gap or the red supergiant phase. The large $\epsilon_{\rm acc}$ close to $\sim 10^{-4}$ during this phase favors a massive BH accretor. The somewhat He-rich nature of SN 2023zkd may be explained by an earlier stripping of the H-rich envelope by Case B (post-main sequence) mass transfer. For low metallicity like observed in SN 2023zkd, the donor could leave behind a sizeable mass of H/He-rich layer that later re-expands to 100s of $R_\odot$ after core He depletion \citep{Gotberg17,Laplace20}. In this case, equation (\ref{eq:Mdot_estimate}) could be an overestimate for the mass transfer history as the envelope carries a smaller fraction of the star's mass.

The two precursors of SN 2023zkd, 2021qqp found may be the tip of the iceberg, since we also expect mergers with lower mass/radii and/or NS companions that would lead to dimmer precursors. The two precursors were at $\sim 200$ Mpc, and at such distances sources even a few mag dimmer would be undetectable by archival wide-field surveys like ZTF. We expect the event rates of such dimmer sources would be understood much better in the LSST era, and we plan to make such predictions in a future study.

For the precursor of Type Ibn SN 2023fyq, \cite{Dong24} and \cite{Tsuna24} proposed mass transfer between a low-mass ($\lesssim 3M_\odot$) He star donor and a NS companion, based on earlier suggestions that these He stars could rapidly expand to several $10$s of $R_\odot$ in the last decade before core-collapse \citep[][see also \citealt{Woosley19,Ercolino25}]{Wu22b}. Inserting the binary properties in equation (\ref{eq:Mdot_estimate}), at the onset of the dynamical phase at $\sim 0.2$ yr before merger, gives
\begin{eqnarray}
    |\dot{M}_*| &\sim& 0.5~M_\odot\ {\rm yr}^{{-1}} \nonumber \\
    &&\times\left(\frac{q}{0.5}\right)\left(\frac{t_m-t}{0.2~ \rm yr}\right)^{\!\!-1.4}\left(\frac{R_*}{50\ R_\odot}\right)^{\!\!0.6}\left(\frac{M_*}{2~M_\odot}\right)^{\!\!0.8},
\end{eqnarray}
which is reasonably consistent with the inferred $|\dot{M}_*|$ in Figure \ref{fig:2023fyq_Mdot_tau}. This equation itself allows a BH companion as well, but it is unlikely that unstable mass transfer would be realized as the He star donor would be less massive than the BH accretor.

The other possibility, based purely on the mass transfer rate, is a merger between a Wolf-Rayet (WR) star and a compact object. While a massive WR star could lead to a comparable mass transfer rate, the dynamical timescales of these stars are much shorter than low-mass He stars, only $t_{\rm dyn}\sim 700$ sec $(R_*/R_\odot)^{3/2}(M_*/5~M_\odot)^{-1/2}$. At the onset of the dynamical instability, the e-folding time of the light curve (and $|\dot{M}_*|$) is of the order of $\sim 10^3$--$10^4$ dynamical times. This appears inconsistent with an unstable mass transfer of a WR donor, though a detailed test requires binary evolution modeling of such systems. Moreover, Type Ibn SNe as a population is better explained by low-mass He star progenitors, based on the observed spectral features \citep{Dessart22,Wang24}, host environments \citep{Sun20,Hong26}, and event rates \citep{Ko25}.

\subsection{Why are these events double peaked?}
\label{sec:double_peaks}
The two Type IIn events discussed here display a clear bump in the optical light curve at $\approx 200$--$300$ days after the first peak \citep[see Figure 5 of][]{Gagliano25b}. For the Type Ibn SN 2023fyq, the bolometric light curve also shows two peaks, with a much shorter interval of $\approx 10$ days \citep[see Figure 5 of][]{Dong24}. In SN 2023fyq, the two peaks were attributed to shock breakout and subsequent cooling emission from the shocked CSM \citep{Dong24}. However, the extremely long intervals in the two Type IIn events seem inconsistent with such a scenario \citep[see e.g.][]{Khatami24}.

Such multiple peaks are rarely seen in Type IIn SNe. Based on a sample of 39 SN IIn with rich post-peak photometry, \cite{Nyholm20} find that a clear double-peaked morphology is seen only in SN iPTF13z \citep{Nyholm17}, constraining the fraction of double-peaked events to $1.4^{+10.6}_{-1.0}\%$. Hence the two Type IIn events\footnote{SN 2022pda, classified as ``Transitional Type Ibn/IIn", is recently reported to also have such characteristics of a long-duration ($\gtrsim 100$ day) precursor and a likely secondary peak \citep{Cai26}.} with long-rising precursors both showing double-peaked light curves are puzzling, and may point to some causal connection with this binary merger scenario.

A hypothesis for SN 2021qqp under the binary merger scenario was raised in \cite{Hiramatsu24}, where the multiple peaks arise from successive episodes of mass ejection prior to merger, for example by eccentric encounters \citep[see also][]{Maeda26}. While this interpretation is plausible, it is unclear how the final binary evolution leading to merger could reproduce the precursor emission, which are both nearly steady and long-rising.

We suggest an alternative possibility, that the two peaks are due to a significant delay between the accretor's plunge-in (end of the ``precursor" phase) and the final explosion. For a donor with a H-rich envelope, the rise to the first peak could be created by the binary leading to merger, as demonstrated by this work. On the other hand, we hypothesize that the second peak is the final explosion (and interaction with the precursor CSM), after the compact object spirals into the envelope. The final outcome is either a common envelope ejection with a surviving binary, or the compact object merging with the donor's He core (or tidally disrupting it before that). 

The second ``explosion" speculated here would have energy of the order of the envelope's binding energy for the former case of envelope ejection, and likely stronger energies for the latter case of cataclysmic disruption. In either case, we expect the delay to be roughly several dynamical timescales of the donor, which could be 100s of days for inflated donors like invoked for SN 2021qqp and 2023zkd. Therefore the observed interval between the two peaks is more naturally explained by this scenario. This also implies that only a single peak may be seen if the dynamical timescale is much shorter than the diffusion timescale of the SN, which could occur for compact donors (e.g. main-sequence or WR stars) with $t_{\rm dyn}$ of hours or less.

(Radiation) hydrodynamical simulations would lead to more quantitative predictions, which we plan to explore in future studies. Nevertheless, a clear observational prediction from the our hypothesis of delayed explosion is that we expect to only observe up to two clear peaks. A larger sample of these events with long-term photometry, spanning years after the first peak, could lead to deeper understanding of the binary progenitors of these events.

\section{Conclusion}
\label{sec:conclusion}
We developed a semi-analytical light curve model to characterize long-duration transients from unstable mass transfer onto a compact object companion. Such transients have recently been considered to be responsible for some precursors of interacting SNe with long rises spanning years. By inferring the mass transfer history of these systems from the precursor emission, our work is the first step towards understanding the binary progenitors of these events, and many similar events expected to be discovered in the LSST era.

As a demonstration, our model was applied to the long-duration precursors of two SNe IIn (SN 2023zkd and SN 2021qqp), and one SN Ibn (SN 2023fyq). Based on the inferred mass transfer history and previous suggestions, we favor evolved donors with a large radius ($\gtrsim 100~R_\odot$) and BH companions for the two SN IIn events, and a low-mass He star donor with likely a NS companion for SN 2023fyq. We believe the observed SN IIn events are only the tip of the iceberg, and there would be more numerous dimmer events with lower-mass/radius donors and/or NS companions. This prediction would be testable in the era of LSST, as we accumulate a deeper, statistical sample of these precursor events.

We conclude with a few directions for future work. Firstly, our work has been limited to an inverse problem approach, where we have inferred the binary properties from the light curves and demonstrated the feasibility of binary models to reproduce these events. The important next step is the forward modeling, i.e. predicting the mass transfer histories themselves and the precursor emission at a population level, as well as understanding the observational biases when interpreting the population from observations. Such predictions would be key in the LSST era when we expect to find tens to hundreds of these events, and we plan to explore this in future work.

Secondly, our model has focused on the precursor phase, where the evolution of the mass transfer and accretion are relatively well understood. We currently lack predictions for the rich phenomenology of these events at later phases, such as a sharp peak and a secondary bump. This is also the phase approaching the merger and the common envelope phase, where the binary evolution becomes less tractable analytically. Numerical simulation including accretion physics, possibly in a sub-grid manner using the model here, is a feasible next step towards understanding this phase.

Our model also has strong physical connections to LRNe, which are also great LSST targets and are more observationally robust to be stellar merger events \citep[e.g.,][]{Tylenda11}. There are also several existing LRNe with long-rising precursors detected \citep[e.g.,][]{Tylenda11,Blagorodnova17,Pastorello23}, that encourage constructing a semi-analytical model fitting framework like this work. Yet there are three complexities when applying this model: (i) The inflated donor would be a significant contribution to the earliest emission and influence our precursor fits, (ii) The ``accretion power" in the currect model would be an uncertain fraction of the luminosity dissipated at the accretor's surface ($GM_{\rm acc}\dot{M}_*/R_{\rm acc}$), which potentially could be calibrated from numerical simulations, and (iii) Our model assumes radiation pressure is dominant, and gas pressure could be dominant for dimmer/cooler parts of the parameter space \citep[e.g.,][]{Pejcha16a}. We encourage future studies on constructing such models for LRNe incorporating these effects.

\begin{acknowledgments}
We thank Yize Dong, Alex Gagliano, and Daichi Hiramatsu for providing the light curve data of the three SN precursors analyzed in this work, and for discussions. We also thank Jim Fuller, Wenbin Lu and Samuel Feyan for discussions on the model. D. T. is supported by Harvard University through the Institute for Theory and Computation Fellowship. The Villar Astro Time Lab acknowledges support through the David and Lucile Packard Foundation, the Research Corporation for Scientific Advancement (through a Cottrell Fellowship), the National Science Foundation under AST-2433718, AST-2407922 and AST-2406110, as well as an Aramont Fellowship for Emerging Science Research. This work is supported by the National Science Foundation under Cooperative Agreement PHY-2019786 (the NSF AI Institute for Artificial Intelligence and Fundamental Interactions).  
\end{acknowledgments}

\begin{appendix}
\section{Corner plot for each event}

\begin{figure}
    \centering
    \includegraphics[width=\linewidth]{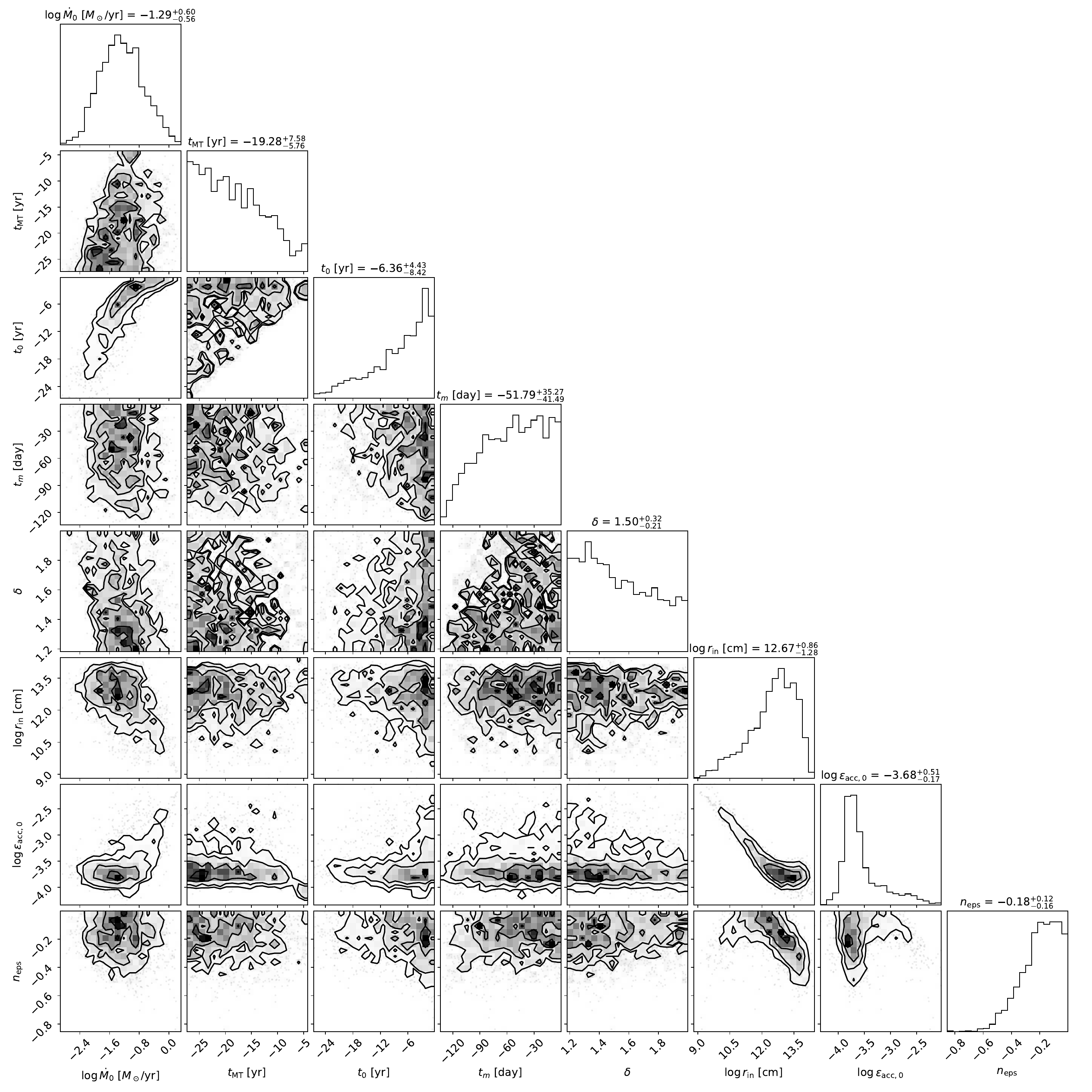} 
    \caption{Parameter estimation for SN 2023zkd.}
    \label{fig:2023zkd_corner}
\end{figure}

\begin{figure}
    \centering
    \includegraphics[width=\linewidth]{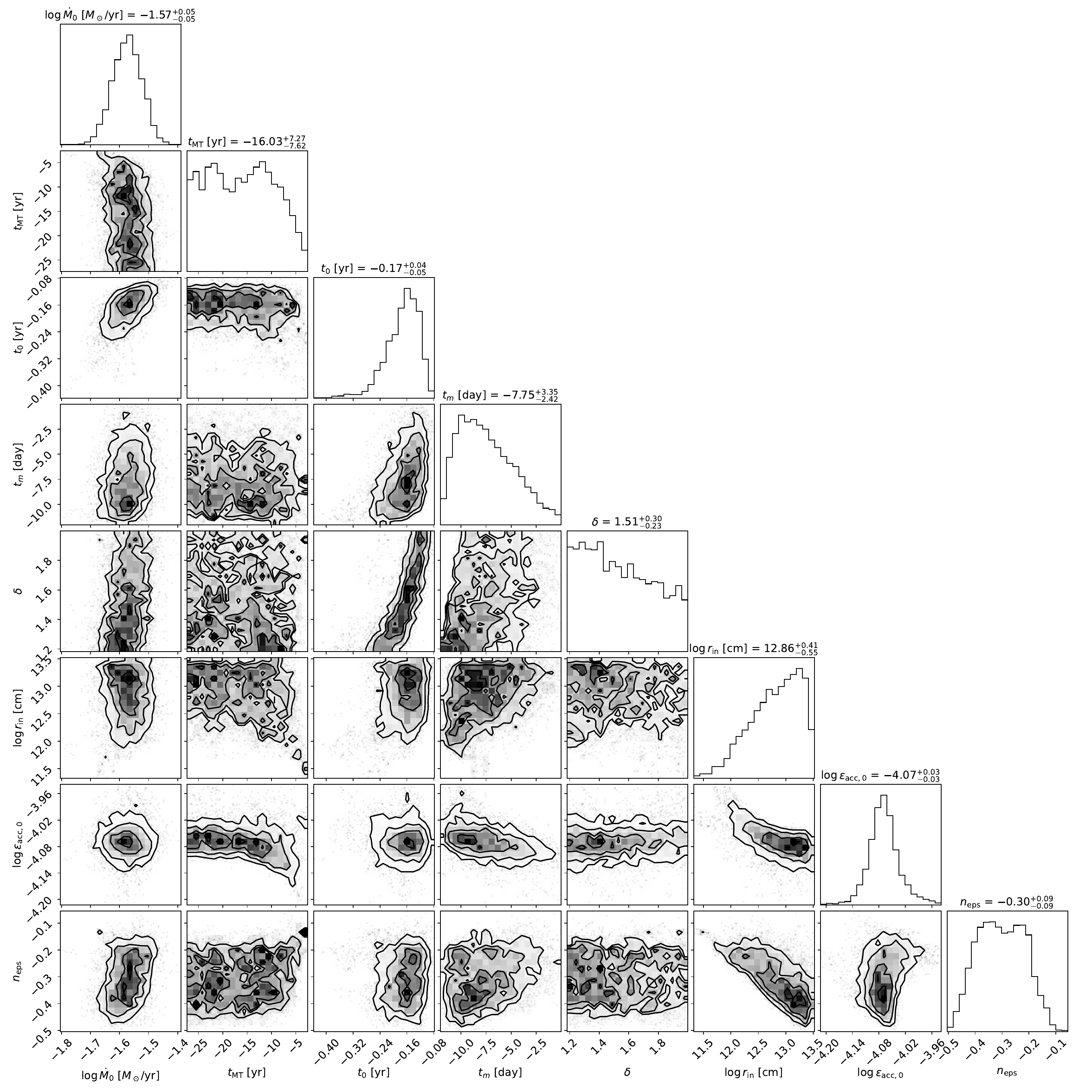} 
    \caption{Parameter estimation for SN 2023fyq.}
    \label{fig:2023fyq_corner}
\end{figure}

\begin{figure}
     \centering
    \includegraphics[width=\linewidth]{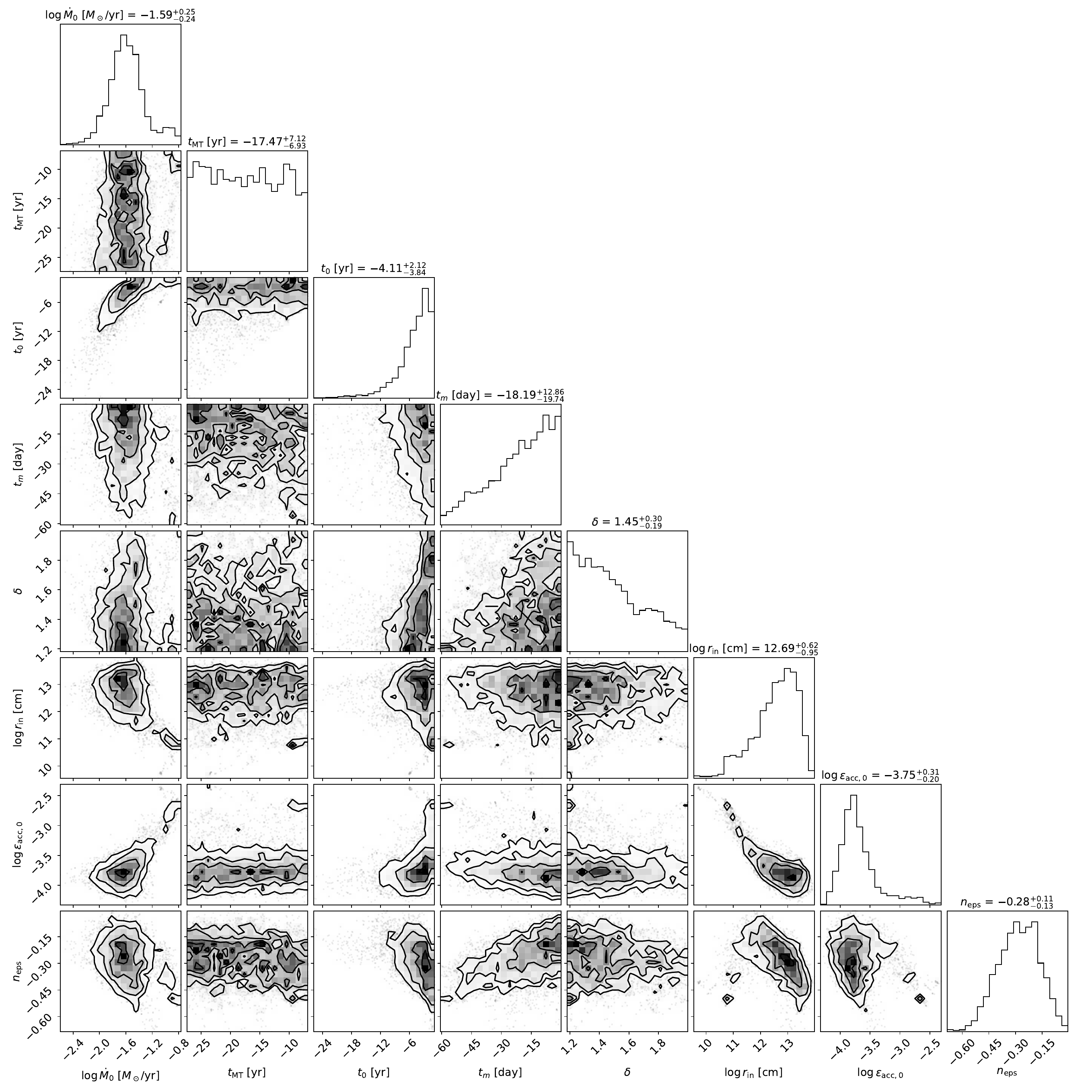} 
    \caption{Parameter estimation for SN 2021qqp.}
     \label{fig:2021qqp_corner}
 \end{figure}

\end{appendix}

\bibliography{references}
\bibliographystyle{aasjournal}

\end{document}